\documentclass[aps,prd,twocolumn,amsmath,showpacs,amssymb,superscriptaddress,nofootinbib,preprintnumbers]{revtex4-1}
\usepackage{graphicx}
\usepackage{bm}
\usepackage{amssymb,amsmath}
\usepackage{mathrsfs}
\usepackage{latexsym}
\usepackage{mathtools} 
\usepackage{color}
\usepackage{float}
\usepackage[normalem]{ulem} 
\usepackage{dcolumn}
\usepackage[usenames,dvipsnames]{xcolor}
\definecolor{keynoteBlue}{HTML}{0365C0}
\usepackage[colorlinks=true,linkcolor=red,citecolor=purple, urlcolor=keynoteBlue!60!black]{hyperref}

\definecolor{pyBlue}{RGB}{31, 119, 180}
\definecolor{pyRed}{RGB}{214, 39, 40}
\definecolor{pyGreen}{RGB}{44, 160, 44}
\definecolor{pyBlue2}{RGB}{0, 111, 237}
\definecolor{pyRed2}{RGB}{224, 52, 36}

\newcommand\myshade{80}
\hypersetup{
  linkcolor  = pyRed!\myshade!black,
  citecolor  = pyRed!\myshade!black,
  urlcolor   =   keynoteBlue!60!black,
  colorlinks = true
}

\allowdisplaybreaks

\newcommand{\JHU}{\affiliation{William H. Miller III Department of Physics and Astronomy, Johns Hopkins University, Baltimore, Maryland 21218, USA}}
\newcommand{\Austin}{\affiliation{The Weinberg Institute, University of Texas at Austin, Austin, TX 78712, USA}}
\newcommand{\IAS}{\affiliation{School of Natural Sciences, Institute for Advanced Study, Princeton, NJ 08540, USA}}
\newcommand{\LIGOMIT}{\affiliation{LIGO Laboratory, MIT, Cambridge, MA 02139, USA}}
\newcommand{\Stockholm}{\affiliation{The Oskar Klein Centre, Department of Physics, Stockholm University, AlbaNova, SE-106 91 Stockholm, Sweden}}
\newcommand{\UdeM}{\affiliation{Département de Physique, Université de Montréal, 1375 Avenue Thérèse-Lavoie-Roux, Montréal, QC H2V 0B3, Canada}} 
\newcommand{\Mila}{\affiliation{Mila -- Quebec AI Institute, 6666 St-Urbain, \#200, Montreal, QC, H2S 3H1}} 
\newcommand{\NORDITA}{\affiliation{Nordic Institute for Theoretical Physics (NORDITA), 106 91 Stockholm, Sweden}}

\def\beq{\begin{equation}}
\def\eeq{\end{equation}}

\newcommand{\kappatwo}{\kappa^{\rm 2PN}_{\rm eff}}
\newcommand{\kappathree}{\kappa^{\rm 3PN}_{\rm eff}}

\begin{document}

\title{Dimensionally Reduced Waveforms for Spin-Induced Quadrupole Searches}

\author{Horng Sheng Chia}
\email[Electronic address: ]{hschia@ias.edu}
\IAS
\author{Thomas D.~P.~Edwards}
\email[Electronic address: ]{thomas.edwards@fysik.su.se}
\Stockholm \NORDITA\JHU
\author{Richard N.~George}
\email[Electronic address: ]{rngeorge@utexas.edu}
\Austin
\author{Aaron Zimmerman}
\email[Electronic address: ]{aaron.zimmerman@austin.utexas.edu}
\Austin
\author{Adam Coogan}
\UdeM \Mila
\author{Katherine Freese}
\Austin \Stockholm \NORDITA
\author{Cody Messick}
\LIGOMIT
\author{Christian N. Setzer}
\Stockholm

\preprint{UTWI-13-2022, LIGO-P2200325}

\date{\today}

\begin{abstract}
\noindent We present highly accurate, dimensionally-reduced gravitational waveforms for binary inspirals whose components have large spin-induced quadrupole moments. 
The spin-induced quadrupole of a body first appears in the phase of a waveform at the early inspiral stage of the binary coalescence, making it a relatively clean probe of the internal structure of the body.
However, for objects with large quadrupolar deviations from Kerr, searches using binary black hole (BBH) models would be ineffective. 
In order to perform a computationally-feasible search, we present two dimensionally-reduced models which are derived from the original six-dimensional post-Newtonian waveform for such systems. 
Our dimensional reduction method is guided by power counting in the post-Newtonian expansion, suitable reparameterizations of the source physics, and truncating
terms in the phase that are small in most physically well-motivated regions of parameter space.
In addition, we note that large quadrupolar deviations cause the frequency at which a binary system reaches its minimum binding energy to be reduced substantially.
This minimum signals the end of the inspiral regime and provides a natural cutoff for the PN waveform. 
We provide accurate analytic estimates for these frequency cutoffs.
Finally, we perform injection studies to test the effectualness of the dimensionally reduced waveforms. We find that over $80\%$ of the injections have an effectualness of $\varepsilon > 0.999$, significantly higher than is typically required for standard BBH banks, for systems with component spins of $|\chi_i| \lesssim 0.6$ and dimensionless quadrupole of $\kappa_i \lesssim 10^3$.
Importantly, these waveforms represent an essential first step towards enabling an effective search for astrophysical objects with large quadrupoles. 

\end{abstract}

\maketitle
 
\section{Introduction}

The direct detection of gravitational waves from coalescing compact objects has provided a fundamentally new way to observe the Universe. 
Ever since the first detection of a binary black hole (BBH) in 2015~\cite{Abbott:2016blz}, the Advanced LIGO~\cite{LIGOScientific:2014pky} and Virgo~\cite{VIRGO:2014yos} observatories have detected merging black holes and neutron stars at an increasing rate, with over 100 binary systems observed to date~\cite{LIGOScientific:2018mvr, Venumadhav:2019tad, Venumadhav:2019lyq, Nitz:2018imz, Nitz:2020oeq, LIGOScientific:2020ibl, Nitz:2021uxj, LIGOScientific:2021djp, Nitz:2021zwj, Olsen:2022pin}.
This large influx of data, along with the exciting prospect of next-generation observatories~\cite{Adhikari:2019zpy, Somiya:2011np, Unnikrishnan:2013qwa, Sathyaprakash:2012jk, Reitze:2019iox, 2017arXiv170200786A, Luo:2015ght, Guo:2018npi} naturally raises many questions: in addition to black holes and neutron stars, can gravitational-wave observations shed light on the existence of new types of compact objects? 
If new objects do exist in Nature, do we have the necessary methods, such as the right template waveforms, to search for them?


One of the most promising ways of detecting the existence of new types of compact objects is through their spin-induced quadrupole moments~\cite{Hansen:1974zz, Thorne:1980ru, Poisson:1997ha}. 
The spin-induced quadrupole of a self-gravitating body is sourced by the rotational motion of the body, and first appears in the phase of binary waveforms at the second post-Newtonian (PN) order~\cite{Poisson:1997ha}. 
This relatively low-order 2PN effect can contribute sizably to the phase evolution of a gravitational waveform. 
In fact, from a purely PN power counting perspective, the quadrupole moment is the dominant of all finite-size effects: other contributions such as the tidal deformability~\cite{Flanagan:2007ix, Binnington:2009bb, Damour:2009vw, Chia:2020yla, Charalambous:2021mea} and tidal dissipation~\cite{Poisson:1994yf, Tagoshi:1997jy, Chia:2020yla, Goldberger:2020fot, Charalambous:2021mea, LeTiec:2020spy} of a rotating body first appear in waveform phases at 5PN and 2.5PN order, respectively. 
Interestingly, the (dimensionless) spin-induced quadrupoles of Kerr black holes and neutron stars are comparable in size~\cite{Hansen:1974zz, Thorne:1980ru, Laarakkers:1997hb, Pappas:2012ns} -- a fact that allows us to conveniently use BBH waveforms to search for binary neutron stars~\cite{LIGOScientific:2018mvr, Venumadhav:2019tad, Venumadhav:2019lyq, Nitz:2018imz, Nitz:2020oeq, LIGOScientific:2020ibl, Nitz:2021uxj}. 

However, for more speculative types of compact objects, which exist in many scenarios of physics beyond the Standard Model, the spin-induced quadrupoles can easily be orders of magnitude larger than those of black holes and neutron stars~\cite{Ryan:1996nk, Herdeiro:2014goa, Baumann:2018vus, Baumann:2019ztm}.
In this case, they may be missed by matched-filtering searches which use BBH template waveforms~\cite{Chia:2020psj}. 
This is especially true for low mass systems, with total mass $\lesssim 10 M_\odot$, which produce long inspiral signals in current ground-based detectors, and is the focus of this work.
For long signals, even small inaccuracies in the predicted phase evolution of template waveforms can result in significant dephasing with a true gravitational wave signal.


In order to detect binaries whose components have large spin-induced quadrupole moments, one would need to construct a bank of templates using a waveform model that accounts for these moments.
Fortunately, a frequency-domain inspiral-only PN waveform model that accommodates arbitrary values of spin-induced quadrupole moments has been developed~\cite{Krishnendu:2017shb}. 
This waveform model has been used to measure the spin-induced quadrupoles of the catalog of detected binary systems~\cite{Krishnendu:2019tjp, LIGOScientific:2020tif,LIGOScientific:2021sio}.
These studies constrained deviations to the sum of the quadrupoles of the two bodies, which provides the dominant effect on the phase evolution of the signal. 
Since those binary signals were detected using BBH template waveforms, the measured quadrupolar deviations from Kerr are expected to be small, and this is indeed the case (up to measurement uncertainties).


In principle, the model of~\cite{Krishnendu:2017shb} can be used directly in constructing template banks to search for binaries with large spin-induced quadrupoles. 
However, doing so in practice is challenging: since the number of dimensions of the intrinsic parameter space of the PN waveform is large, the resulting template bank could easily be so large that matched filtering searches are computationally prohibitive.
For comparison, current searches using BBH templates templates typically use waveforms for quasi-circular binaries, where the spins of the components are aligned with the orbital angular momentum (precession effects are neglected).
This results in four intrinsic parameters, the two masses of the binary components $\{m_1, m_2\}$ and the two dimensionless spin components $\{\chi_1, \chi_2\}$, with $-1 \leq \chi_{1, 2} \leq 1$.
Such template banks typically contain $\mathcal{O}(10^5 \text{--} 10^6)$ templates~\cite{Usman:2015kfa, Messick:2016aqy, Roulet:2019hzy}. 
On the other hand, the PN waveform for arbitrary spin-induced quadrupoles~\cite{Krishnendu:2017shb} has a six-dimensional intrinsic parameter space, with two additional parameters describing the magnitudes of the spin-induced quadrupoles, $\{\kappa_1, \kappa_2\}$. 
These spin-induced quadrupole parameters are only bounded from below by the Kerr black hole's value but could otherwise take arbitrarily large values, $\kappa_{1,2} \geq 1$.
Absent a specific compact object in mind, we could in principle extend our search coverage over a wide range of quadrupole parameters, as much as computational resources permit. 
These factors rapidly increase the size of the template bank, potentially rendering matched-filtering searches of these new types of signals computationally unfeasible.


This issue can be alleviated by constructing a lower dimensional model which still provides a sufficiently good match to the full six dimensional model over the parameter ranges of interest.
Such an approach has been carried out previously, in order to construct a three dimensional BBH template bank to cover the space of four-dimensional BBH templates~\cite{Ajith:2011ec,Ajith:2012mn}.
The strategy is to select a reduced set of intrinsic parameters which cover the most important contributions to the phase evolution, and truncate the remaining contributions to the phase. 


In this work, we begin with the original inspiral-only PN waveform developed in~\cite{Krishnendu:2017shb}, and construct two dimensionally-reduced waveforms, one five-dimensional and the other four-dimensional, which would permit computationally-realistic searches for binary signals with large spin-induced quadrupoles. 
Our dimensional-reduction method involves a combination of strategies.
Firstly, guided by power counting in the PN expansion, we choose to only truncate phase terms that appear at high PN orders whenever necessary.
Secondly, we map the original PN parameters to reduced spin and effective quadrupole parameters to avoid any truncations in the phase at low PN orders. 
As such, both of our models are exact up to 2PN order in the phase. 
Finally, building on the approach of~\cite{Ajith:2011ec,Ajith:2012mn}, we truncate terms that are not captured by our effective quadrupole parameters by setting the anti-symmetric spin combination, $\chi_a = (\chi_1 - \chi_2)/2$, to vanish in our models. 
The truncation error only becomes sizable when \textit{i)} the binary spin orientations are anti-aligned with each other and \textit{ii)} the component spin magnitudes are large. 
While the former depends on the formation mechanism of the binary population and is \textit{a priori} hard to model, the latter is unlikely when both binary components are exotic compact objects because the surfaces of most astrophysical objects are bounded by the Keplerian limit and the spins are far from extremality. 
Nevertheless, if one of the binary constituents is a black hole near extremality, the truncation errors of our reduced waveforms would be large. 


We also present an effectualness study for these reduced waveforms, showing that they attain an effectualness of $\varepsilon > 0.999$, significantly higher than is typically demanded of standard BBH banks, over a wide range of parameter space. 
Specifically, these reduced models are effective at searching for binary components with $|\chi| \lesssim 0.6$ and $\kappa_i \lesssim 10^3$, and are especially reliable when the object with large spin-induced quadrupole is the heavier counterpart. 
Furthermore, we shall show that the effectualness of the four-dimensional waveform degrades only slightly from that of the five-dimensional waveform. 
As such, in practice we are able to restrict ourselves to the four-dimensional waveform in actual searches. 
The template bank construction and the results of the spin-induced quadrupole inspiral-only searches will be reported in companion papers~\cite{Coogan:2022qxs, companion1}.


The rest of this paper is structured as follows: in Sec.~\ref{sec:models} we recap the basics of the PN waveform and describe our method of dimensional reduction to achieve two reduced waveform models. 
In Sec.~\ref{sec:effectualness} we perform a full effectualness analysis and study the implications of using the reduced waveforms on our search sensitivity. 
We conclude and present an outlook in Sec.~\ref{sec:conclusions}. 
Finally,  in Appendix~\ref{sec:EnergyAppx} we present the details of the post-Newtonian binding energy after dimensional reduction, whose minima determines the cutoff frequency of our reduced models.
In this paper, the binary components are denoted $m_i$, where $i=1,2$ denotes each of the binary component, and we use the convention $m_1 \geq m_2$.
The dimensionless spins, which we assume are aligned with the binary's orbital angular momentum, are $\chi_i$, and the dimensionless quadrupole parameters are denoted $\kappa_i$.

\section{Post-Newtonian Waveforms with Spin-Induced Quadrupoles}
\label{sec:models}

The spin-induced quadrupole moments of binary components leave distinctive imprints in the phase of gravitational waveforms. 
In Sec.~\ref{sec:quad}, we summarize the way in which the spin-induced quadrupole of an astrophysical object is parameterized in the literature. 
In Sec.~\ref{sec:original}, we recap the basics of the PN waveform model for the quasi-circular inspiral of objects with generic spin-induced quadrupoles and spins aligned with the orbital angular momentum (neglecting precession).
There we focus on how the source physics affects the phase of the PN waveform. 
The intrinsic parameter space for this model is $6\mathrm{D}$ (component masses, aligned spins, and the spin-induced quadrupole parameters). 
Leveraging the PN power counting, we then dimensionally reduce the $6\mathrm{D}$ waveform to a $5\mathrm{D}$ waveform in Sec.~\ref{sec:5d} through the introduction of two effective quadrupole parameters. 
We show how this $5\mathrm{D}$ waveform can be further dimensionally-reduced to a $4\mathrm{D}$ waveform in Sec.~\ref{sec:4d}. 
In Sec.~\ref{sec:Cutoff} we discuss the cutoff frequency of the waveforms when the binary components have large quadrupole moments.


Our general strategy is to proceed from lower to higher PN orders, reparametrizing the model in order to isolate terms that can be truncated in the dimensional reduction.
We truncate at the highest PN order possible, and in such a way that only terms which are small compared to other terms at the same PN order are ever neglected.
These considerations guide our selection of the lower dimensional parametrizations, and the resulting $5\mathrm{D}$ and $4\mathrm{D}$ waveforms agree with the full $6\mathrm{D}$ model up to 2PN order. 
For the $5\mathrm{D}$ model, the terms involving the spin-induced quadrupole moments also agree with the $6\mathrm{D}$ model up to 3PN order.

\subsection{Spin-Induced Quadrupole Moment} \label{sec:quad}

The rotational motion of a self-gravitating body sources a hierarchy of multipole moments, which affect the shape of the body and the gravitational field in its vicinity~\cite{Hansen:1974zz, Thorne:1980ru}.\footnote{
In this work, we assume that the self-gravitating body does not have any permanent (non-spinning) quadrupoles; in other words,  it is spherically symmetric when it is not spinning.} 
When these objects are part of binary systems, these multipole moments leave subtle imprints in the gravitational waveforms, through which we could indirectly infer the nature of the binary components. 
The dominant spin-induced multipole moment is the axisymmetric mass-type quadrupole moment, $Q$, which first appears in the phase of a waveform at 2PN order~\cite{Poisson:1997ha}. 
The quadrupole is commonly parameterized through the relation~\cite{Poisson:1997ha}
\beq
Q_i \coloneqq - \kappa_i \hskip 1pt m_i^3 \chi_i^2 \, , \label{eqn:kappa_def}
\eeq
which serves to define the dimensionless quadrupole parameter $\kappa_i$.


The relation (\ref{eqn:kappa_def}) implies that the spin-induced quadrupole of a body depends on both the dimensionless parameter $\kappa_i$ and the spin magnitude $\chi_i$. For Kerr black holes, the quadratic-in-spin relation in (\ref{eqn:kappa_def}) is exact, with $\kappa_i = 1$~\cite{Hansen:1974zz, Thorne:1980ru} and the spin magnitudes ranging anywhere between zero and the Kerr bound, $|\chi_i| \leq 1$. 
On the other hand, neutron stars have spins of $|\chi_i| \lesssim 0.6$~\cite{LIGOScientific:2018hze} and, depending on the nuclear equation of state, $\kappa_i$ ranges between $2 - 10$ and  can even have mild dependences on $\chi_i$~\cite{Laarakkers:1997hb, Pappas:2012ns}. 
Since black holes and neutron stars have comparable values of $\kappa_i \chi_i^2$, BBH template waveforms are sufficiently accurate to search for binary neutron stars. 
Nevertheless, for more speculative objects such as superradiant boson clouds~\cite{Arvanitaki:2010sy, Baumann:2019eav, Baryakhtar:2020gao} and boson stars~\cite{Kaup1968, Colpi1986, Liebling:2012fv, Schunck:2003kk}, $\kappa_i \chi_i^2$ can easily be orders of magnitude larger than those of black holes and neutron stars~\cite{Ryan:1996nk, Herdeiro:2014goa, Baumann:2018vus}. 
In these beyond the Standard Model scenarios, $\kappa_i$ could also vary with time and strongly depend on $\chi_i$~\cite{Baumann:2019ztm}. 
The wide range of plausible values for the quadrupole moment of compact objects suggest that BBH templates could miss inspiral signals from more unusual types of binary systems. 
Indeed, it was found~\cite{Chia:2020psj} that the effectualness of BBH template banks significantly degrades for binary systems with $\kappa_i \gtrsim 20$ and moderate to large spins.
Without specifying any model of compact object, we will henceforth treat $\kappa_i$ as a free parameter in this work, with the requirement that $\kappa_i \geq 1$.

\subsection{Six-Dimensional Waveform} \label{sec:original}

As discussed above, our waveform model depends on six intrinsic parameters, which we denote
$ \textbf{p}_{6\mathrm{D}} = \{ m_1, m_2, \chi_1, \chi_2, \kappa_1, \kappa_2 \}$.\footnote{
In principle, the post-Newtonian waveform at 3.5PN order contains two additional free parameters for the spin-induced octupole moments of the binary components, $O_i \coloneqq - \lambda_i m_i^4 \chi^3_i$~\cite{Marsat:2014xea, Krishnendu:2017shb}. 
For simplicity, we set $\lambda_i = 1$ (i.e., the black hole values~\cite{Hansen:1974zz, Thorne:1980ru}) in this work. \label{footnote:octupole}} 
In Fourier space, the dominant quadrupolar gravitational radiation strain is 
\beq
\label{eqn:waveform}
\tilde{h}(f; \textbf{p}_{6\mathrm{D}}) = \mathcal{A}(\textbf{p}_{6D}) f^{-7/6} \hskip 1pt e^{i \psi(f;  \textbf{p}_{6\mathrm{D}})} \hskip 1pt \theta(f_{\rm cut}(\textbf{p}_{6\mathrm{D}}) - f) \, ,
\eeq
where the amplitude $\mathcal{A}$ is independent of $f$ at leading-order, $\psi$ is the waveform phase, $\theta$ is the Heaviside function, and $f_{\rm cut}$ is the cutoff frequency of the inspiral. The cutoff frequency is often taken to be the frequency at the innermost stable circular orbit (ISCO), which corresponds to the minimum of the binary binding energy (see Sec.~\ref{sec:Cutoff} below), though for general compact objects $f_{\rm cut}$ is also determined by the compactness of the binary components~\cite{Chia:2020psj}.

Since achieving phase coherence is the most important goal in template bank searches, we focus on the higher-order PN components of the phase. 
In the stationary phase approximation~\cite{Sathyaprakash:1991mt, Cutler:1994ys}, $\psi$ is
\begin{align}
\psi (f; \textbf{p}_{6\mathrm{D}})  = & \,\, 2 \pi f t_c - \phi_c - \frac{\pi}{4} 
\notag \\
& \,+ \frac{3}{128 \hskip 1pt \nu \hskip 1pt v^5} \Big[ \psi_{\rm NS} (f; m_1, m_2) + \psi_{\rm S} (f; \textbf{p}_{6\mathrm{D}}) \Big] \, , 
\label{eqn:phase}
\end{align}
where $t_c$ is the time of coalescence, $\phi_c$ is the phase of coalescence, $\nu \coloneqq m_1 m_2 / (m_1 + m_2)^2 \leq 1/4$ is the dimensionless reduced mass, $v = (\pi m f)^{1/3}$ is the PN expansion parameter for circular orbits, and $m = m_1 + m_2$ is the total mass.
The component of the phase which does not depend on the spins is denoted $\psi_{\rm NS}$, and the spin-dependent part is $\psi_{\rm S}$. 


Through heroic efforts over the past few decades, the analytic post-Newtonian expression for the phase \eqref{eqn:phase} of an aligned-spin, quasi-circular orbit has been completed up to 3.5PN order through myriad techniques~\cite{Arnowitt:1962hi, Blanchet:2006zz, nrgr, Bern:2019crd}. 
The non-spinning part of the phase up to 3.5PN order is~\cite{Arun:2004hn, Buonanno:2009zt, Mishra:2016whh}
\begin{align}
\psi_{\rm NS} & (f; m_1, m_2)  = \,  1 + \left( \frac{3715}{756} + \frac{55}{9} \nu \right) v^2 - 16 \pi  v^3 \nonumber \\  & +  \left( \frac{15293365}{508032} + \frac{27145}{504} \nu + \frac{3085}{72} \nu^2 \right) v^4 
\nonumber \\ &
+ \pi \left( \frac{38645}{756} - \frac{65}{9} \nu \right) \left[ 1 + 3 \log \left( v \right)  \right] v^5  \label{eqn:psiNS}  \\
& + \bigg[ \left( \frac{11583231236531}{4694215680} - \frac{640}{3} \pi^2 - \frac{6848 }{21}  \gamma_E \right)   
\nonumber \\ &
+ \left( \frac{2255 \pi^2}{12} - \frac{15737765635}{3048192}  \right) \nu  \nonumber  \\ & + \frac{76055}{1728} \nu^2 - \frac{127825}{1296} \nu^3 - \frac{6848 }{21}  \log (4 v)  \bigg] v^6 
\nonumber \\ &
+ \pi \left( \frac{77096675}{254016} + \frac{378515}{1512} \nu - \frac{74045}{756} \nu^2 \right) v^7 \nonumber \, ,
\end{align}
where $\gamma_E$ is the Euler-Mascheroni constant. 
The terms in \eqref{eqn:psiNS} approximate the binary components as point particles and therefore only depend on $m_i$. 
On the other hand, the expression for the spin-dependent phase $\psi_{\rm S}$ up to 3.5PN order, valid for arbitrary values of $\kappa_i$ and $\chi_i$, is more complicated~\cite{Krishnendu:2017shb}. 
It is convenient to organize the terms in $\psi_{\rm S}$ as
\beq
\psi_{\rm S}(f; \textbf{p}_{6\mathrm{D}}) = \sum_{n=3}^7 S_n (\textbf{p}_{6\mathrm{D}}) \hskip 1pt  v^n + S_{5, \log} (\textbf{p}_{6\mathrm{D}}) \hskip 1pt v^5 \log\left( v \right) \, , \label{eqn:spin_phase}
\eeq
where the precise expressions for ${S_n}$ and $S_{5, \log} = 3 S_5$ can be found in~\cite{Krishnendu:2017shb}. 
The central goal of this work is to find good approximations and reparameterizations to (some of) $S_n$ and $S_{5, \log}$ that dimensionally reduce the $6\mathrm{D}$ parameter space to five and four dimensions, in a way that does not significantly reduce the accuracy of the PN waveform phase. 


In both of our 5D and 4D reduced waveforms, the leading-order spin-dependent term in \eqref{eqn:spin_phase} is parameterized in the same manner. 
Specifically, the $S_3$ coefficient can be succinctly parameterized by the \textit{reduced spin}, $\chi$, as~\cite{Kidder:1992fr, Poisson:1995ef, Ajith:2011ec}:
\beq
S_3 =   \frac{113}{3} \chi \, , \qquad  \chi \coloneqq \left( 1 - \frac{76}{113} \nu \right)  \chi_s  + \delta  \chi_a \, , \label{eqn:SO1.5}
\eeq
where $\chi_s = (\chi_1 + \chi_2)/2$ and $\chi_a = (\chi_1 - \chi_2)/2$ are the symmetric and antisymmetric spins, while $\delta = (m_1-m_2)/(m_1+m_2) = {(1-4\nu)}^{1/2}$ is the asymmetric mass ratio. 
This phase term is sourced by the spin-orbit interaction of the binary system~\cite{Barker75, HartleThorne1985, Kidder:1995zr, Kidder:1992fr}, and affects the phase at 1.5PN order. 


It is important to emphasize that $\chi$ is different from the well-known effective spin parameter $\chi_{\rm eff}$~\cite{Racine:2008qv, Ajith:2009bn, Santamaria:2010yb}, which is defined as $\chi_{\rm eff} = \chi_s + \delta \chi_a $. 
While the difference between $\chi$ and $\chi_{\rm eff}$ is small in the extreme-mass-ratio limit $\nu \ll 0$, (e.g.~\cite{Ajith:2009bn}), it can be important for comparable mass systems where a large number of cycles are observed. Since we focus on low-mass binary systems $m \lesssim 10 M_\odot$ in this work~\cite{Chia:2020psj}, 
we adopt $\chi$ as defined in \eqref{eqn:SO1.5} as one of our model parameters, and so avoid any form of approximation at this order.

\subsection{Reduced Five-Dimensional Waveform} \label{sec:5d}

The phase terms \eqref{eqn:psiNS} and \eqref{eqn:SO1.5} already exist in the PN literature. Importantly, those terms are neither approximated nor reparameterized in our reduced models. 
In the rest of this section, we demonstrate how the 2PN and higher-order spin-dependent terms must either be reparameterized or approximated in order to reduce the dimension of parameter space. 
Particularly, we reduce the parameter space from 6D to 5D, $\textbf{p}_{6\mathrm{D}} = \{ m_1, m_2, \chi_1, \chi_2, \kappa_1, \kappa_2 \} \to  \textbf{p}_{5\mathrm{D}} = \{ m_1, m_2, \chi, \kappa_{\rm eff}^{\rm 2PN}, \kappa_{\rm eff}^{\rm 3PN} \}$, through the introduction of two effective quadrupole parameters.


The 2PN order is the lowest order at which our parameterization of the phase differs from that of the original waveform~\cite{ Krishnendu:2017shb}, as it is the dominant order at which the spin-induced quadrupole first appears in the phases of waveforms~\cite{Poisson:1997ha}. 
The coefficient $S_4$ in~\eqref{eqn:spin_phase} for a general compact binary is~\cite{ Krishnendu:2017shb}
\begin{align}
S_4 = & - \frac{5}{8}  \Big[ 1 + 156 \nu +  80 \, \delta \, \kappa_a + 80 \left( 1 - 2\nu \right) \kappa_s\Big] \chi_s^2 
\nonumber \\ &
 - \frac{5}{8}  \Big[ 1 - 160 \nu +  80 \, \delta \, \kappa_a + 80 \left( 1 - 2\nu \right) \kappa_s\Big] \chi_a^2 \nonumber \\
&  - \frac{5}{4} \Big[ \delta + 80 \, \delta \, \kappa_s +80 (1-2\nu) \kappa_a \Big] \chi_s \chi_a  \, ,  
\label{eqn:N4S}
\end{align}
where $\kappa_s \coloneqq (\kappa_1 + \kappa_2)/2$ and $\kappa_a \coloneqq (\kappa_1 - \kappa_2)/2$. 
By making the substitution $\{ \chi_s, \chi_a \} \to \{ \chi_1, \chi_2\}$ and reorganizing the terms, one can identify three physical effects that contribute to \eqref{eqn:N4S}~\cite{Barker75, Kidder:1992fr, Kidder:1995zr, Poisson:1995ef, Arun:2008kb, Mikoczi:2005dn}: \textit{i)} the mutual spin-spin interaction, \textit{ii)} the spin-induced quadrupole moment \eqref{eqn:kappa_def}, and \textit{iii)} the point-particle self-spin effect, which arises due to the current-quadrupole interaction in the radiative sector. 
In order to parameterize \eqref{eqn:N4S} in terms of the reduced spin $\chi$, as defined in \eqref{eqn:SO1.5}, it is convenient to introduce the \textit{2PN effective quadrupole parameter}
\begin{align}
\label{eqn:kappa_eff_2PN}
\kappatwo \coloneqq & \, \left(\frac{m_1}{m}\right)^2 \left(\kappa_1 - \frac{11065m_1 +7265m_2}{7500m_1 +11300 m_2} \right)  \chi_1^2 
\nonumber \\ &
 + \, (1 \leftrightarrow 2) \, ,
\end{align}
such that (\ref{eqn:N4S}) can be succinctly rewritten as
\beq
\begin{aligned}
S_4  = - 50 \, \kappa^{\rm 2PN}_{\rm eff} -\frac{395}{8} \left( \frac{75}{113} \hskip1 pt \delta^2 + \frac{35344}{12769} \hskip 1pt \nu \right)^{-1} \chi^2 \, . \label{eqn:N4SnuKappa}
\end{aligned}
\eeq
Note that (\ref{eqn:N4SnuKappa}) is the unique way in which all of the mutual spin-spin interaction terms are absorbed by $\chi^2$, while (approximately) preserving the $\kappa_i \chi_i^2$  scaling of \eqref{eqn:kappa_def} in $\kappa_{\rm eff}^{\rm 2PN}$. 


Crucially, $\kappa_{\rm eff}^{\rm 2PN}$ is defined such that neither approximations nor truncations are necessary at this PN order. As a result, in addition to the $(m_i/m)^2 \kappa_i \chi_i^2$ terms, which arise directly from (\ref{eqn:kappa_def}), there are residual $\kappa_i$-independent terms in (\ref{eqn:kappa_eff_2PN}) which absorb any point-particle self-spin effects that cannot be captured by the $\chi^2$ term. 
Interestingly, since the $ (11065m_1 +7265m_2)/(7500m_1 +11300m_2)$ factor exceeds unity when $m_2 / m_1 \lesssim 0.9$, $\kappa_{\rm eff}^{\rm 2PN}$ can become negative for unequal-mass binary systems in which the heavier component is a black hole, $\kappa_1=1$. 
On the other hand, $\kappa_{\rm eff}$ would be positive for all binary configurations if the heavier object has $\kappa_1 > 1.475$, which as we discussed in \S\ref{sec:quad} essentially applies to all other types of compact objects.

\begin{figure}[t!]
\includegraphics[width=0.98\columnwidth]{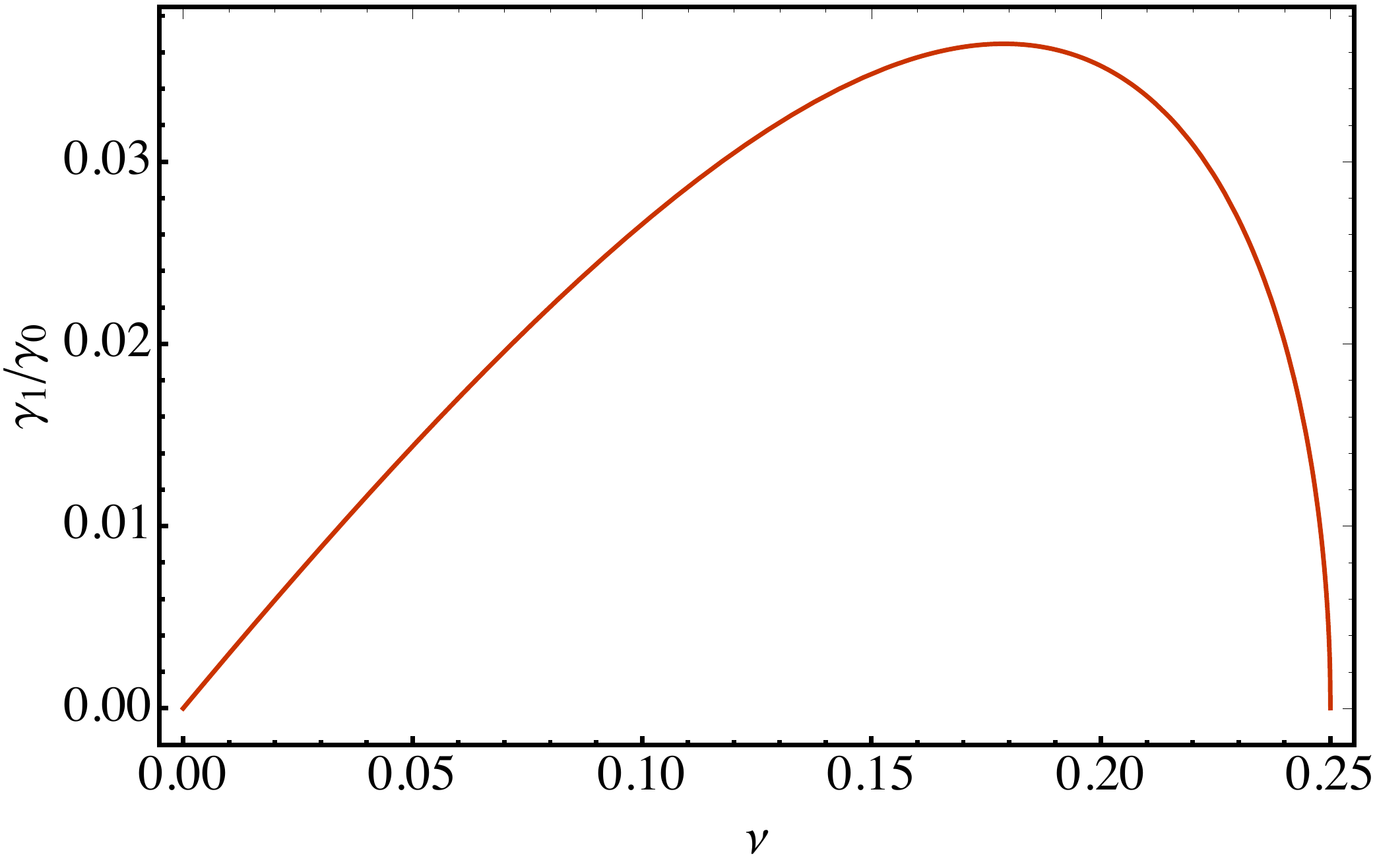} 
\caption{The ratio of the mass-dependent prefactors $\gamma_1$ and $\gamma_0$ arising in $S_5$. 
The truncated term is $\Delta S_5 = \gamma_1 \chi_a$, which vanishes in the equal mass and extreme mass ratio limits, and the ratio of the coefficients is always small.
The maximum phase error occurs at the extremum of $\gamma_1$, which is at $\nu = 0.176$, which corresponds to $m_2 / m_1 \approx 0.3$. 
}
\label{fig:DeltaS5}
\end{figure}

Thus far, our reduced model only consists of a reparameterization of the original PN terms, and we have not made any approximations to the phase. However, this is no longer possible at 2.5PN order and higher, as the four parameters $\{ m_1, m_2, \chi, \kappa_{\rm eff}^{\rm 2PN} \}$ can no longer fully absorb all of the phase terms. 
For instance, the coefficient $S_5$ in \eqref{eqn:spin_phase}, which arises due to higher-order corrections to the spin-orbit interaction~\cite{Faye:2006gx, Blanchet:2006gy}, is
\beq
S_5 = \gamma_0 \hskip 1pt \chi + \Delta S_5 \, , \qquad \Delta S_5 = \gamma_1 \hskip 1pt \chi_a \, ,\label{eqn:S5truncation} 
\eeq
where we have separated terms that depend on $\chi$ and $\chi_a$, and the mass-dependent prefactors are 
\beq
\begin{aligned}
\gamma_0 & = - \left(  \frac{113}{113-76\nu} \right) \left( \frac{732985}{2268} - \frac{24260}{81} \nu - \frac{340}{9} \nu^2  \right) \, , \\
\gamma_1 & =  - \left( \frac{113}{113 - 76 \nu }\right) \left( \frac{2086535}{21357} + \frac{9260  }{339} \nu \right) \delta \hskip 1pt \nu \, .
\end{aligned}
\eeq
Since $\chi$ cannot capture all of the terms in (\ref{eqn:S5truncation}), we choose to truncate $S_5$ in our reduced model by ignoring $\Delta S_5$, effectively setting $\chi_a = 0$~\cite{Ajith:2011ec}. 
Since the 2.5PN logarithmic term is given by $S_{5, \log} = 3 S_5$, we perform the same $\chi_a=0$ truncation for $S_{5, \log}$. 
Interestingly, the phase error incurred by these truncations is usually small, as the ratio $\gamma_1/\gamma_0$ is much smaller than unity over all mass ratios, see Fig.~\ref{fig:DeltaS5}. 
Furthermore, $\gamma_1$ vanishes in the equal-mass and extreme-mass-ratio limits, $\delta \to 0$ and $\nu \to 0$. 
The term neglected in \eqref{eqn:S5truncation} will be largest only when the spins are anti-aligned with each other and significantly different in magnitude, which is an astrophysically unfavourable configuration.
Finally, since \eqref{eqn:S5truncation} is independent of $\kappa_i$, this small truncation error is kept fixed for all types of compact objects.

At 3PN order, new combinations of $\kappa_i$ appear that are linearly independent of the those combinations appearing in the 2PN term.
Thus we must introduce another effective quadrupole parameter to avoid any truncation at this order.
The 3PN spin-dependent phase receives contributions from higher-order corrections to the quadratic-in-spin effects found in \eqref{eqn:N4S}~\cite{Bohe:2015ana, Mishra:2016whh, Krishnendu:2017shb}. 
Using the $\{ m_1, m_2, \chi, \kappa_{\rm eff}^{\rm 2PN} \}$ parameters, we can reorganize $S_6$ as follows:
\begin{align}
S_6 = & \frac{5}{84}  \left( 15609 - 4186 \nu \right)\kappa^{\rm 2PN}_{\rm eff} + \rho_0 \chi  + \rho_1 \chi^2 
\nonumber \\ &
- \frac{2215 \nu^2}{6} \kappa^{\rm 3PN}_{\rm eff} \, , 
\label{eqn:S6param}
\end{align}
whereby we introduce a new \textit{3PN effective quadrupole parameter}
\begin{align}
\kappa^{\rm 3PN}_{\rm eff}  \coloneqq  \kappa_1 \chi_1^2  +\kappa_2 \chi_2^2     -\frac{6}{2215 \nu^2 } \left( \rho_2 \chi_a + \rho_3 \chi_a \chi + \rho_4 \chi_a^2 \right) \, .  \label{eqn:kappa_eff_3PN} 
\end{align}
The mass-dependent $\rho-$coefficients are
\begin{align}
\rho_0 & = \frac{113 \hskip 1pt \pi }{113 - 76  \hskip 1pt  \nu } \left(\frac{2270}{3} - 520  \hskip 1pt  \nu  \right) \, , \nonumber\\[3pt]
\rho_1 & = \left( \frac{113  }{113 - 76  \hskip 1pt  \nu }\right)^2\frac{1}{8475 + 1444  \hskip 1pt  \nu} \,  \nonumber \\[3pt]
& \hskip 10pt \times \bigg(\frac{4008512167}{672} -\frac{10713456163  }{504} \nu  +\frac{2094712507 }{126}\nu^2  
\nonumber \\ &
\hskip 10pt - \frac{1545080 }{9} \nu ^3 \bigg) \, , \nonumber \\[3pt]
\rho_2 & = \frac{3760  \hskip 1pt \pi 
    }{339-228 \nu } \hskip 1pt \delta  \hskip 1pt  \nu \, , 
    \nonumber \\
    \rho_3 & = \frac{113   \left( 12198367-16731218  \hskip 1pt  \nu \right) }{84 \left( 113-76  \hskip 1pt  \nu \right)^2} \hskip 1pt   \delta \hskip 1pt \nu \, , \nonumber \\[3pt]
    \rho_4 & = \frac{ 109852619  - 341495546  \hskip 1pt  \nu - 89556880  \hskip 1pt  \nu^2 }{21 (113-76  \hskip 1pt  \nu )^2} \hskip 1pt \nu ^2 \, . 
\label{eqn:rho_constants}
\end{align}
By this definition, we absorb all the of the $\kappa_i$-dependent and spin-dependent terms that cannot be parameterized by $\{ m_1, m_2, \chi, \kappa_{\rm eff}^{\rm 2PN} \}$ into $\kappa_{\rm eff}^{\rm 3PN}$.


Empirically, we find that if we neglect the $\kappa_{\rm eff}^{\rm 3PN}$-dependent term in \eqref{eqn:S6param} the phase error can greatly exceed $\mathcal{O}(1)$, especially for large values of $\kappa_i$.
This would appear to make further truncation to 4D unacceptable when searching for compact binaries with large spin-induced quadrupole moments.
However, in Sec.~\ref{sec:4d} we find that we can partially account for this term with our first four parameters.
In Sec.~\ref{sec:effectualness} we will show that the effectualness of this 4D model is comparable to the 5D model. 
More remarkably, the effectualness of the 4D model to the full 6D signal waveform is close to unity over a vast range of the parameter space.


Before turning to the 4D model, we complete our 5D model by identifying the appropriate truncations at 3.5PN order.
At this order, the spin-dependent phase contains highly non-trivial linear-in-spin and cubic-in-spin terms. 
In addition to the higher-order corrections to the spin-orbit coupling and spin-induced quadrupole effect, the leading-order spin-induced current-type octupole moment first appears at this PN order~\cite{Bohe:2012mr, Marsat:2014xea, Bohe:2015ana}.
For notational convenience, we decompose $S_7$ into terms that scale linearly and cubically in spins: 
\beq
S_7 = S_7^{\rm lin} + S_7^{{\rm cub}, pp}  + S_7^{\rm cub, {\kappa}} + S_7^{\rm cub, {\lambda}}  \, , \label{eqn:S7}
\eeq
where the cubic-in-spin term is further separated into the point-particle $S_7^{{\rm cub}, pp}$, quadrupole $S_7^{\rm cub, {\kappa}}$, and octupole $S_7^{\rm cub, {\lambda}}$ terms. 
Like the 2.5PN term, \eqref{eqn:S7} contains spin dependence which is not captured by the reduced spin $\chi$, as well as new combinations of $\kappa_i$, and so we are forced to approximate $S_7$ in order to reduce the dimension of parameter space. 


Our strategy is guided by the approach used for the 2.5PN order term \eqref{eqn:S5truncation}, in which we express all of the terms in \eqref{eqn:S7} in the spin variables $\chi$ and $\chi_a$ and then truncate the phase by taking $\chi_a=0$. 
For instance, $S_7^{\rm lin} $ can be written as
\begin{align}
S_7^{\rm lin}  = &  \left( -\frac{25150083775}{3048192} + \frac{10566655595}{762048} \hskip 1pt \nu  - \frac{1042165}{3024} \hskip 1pt \nu^2  
\right.
\nonumber \\ & 
\left. + \frac{5345}{36} \hskip 1pt \nu^3 \right)  
\left( \frac{113}{113-76\hskip 1pt \nu} \right)  \chi  + \Delta S_7^{\rm lin} \,, \\[2pt]
\Delta S_7^{\rm lin} = &    \left( \frac{5 }{113 - 76 \nu} \right) \left( \frac{5575530433}{63504} +\frac{15249235 }{252}\nu  
\right. \nonumber \\ & \left.
+ \frac{24542}{9} \nu ^2 \right)  \delta \hskip 1pt  \nu \hskip 1pt \chi_a \, .  
\label{eqn:psiS7_linear}
\end{align}
and so we neglect $\Delta S_7^{\rm lin} \propto \chi_a$ in our reduced model. For the octupolar cubic-in-spin term, we set octupolar parameters to the black hole values $\lambda_i =1$ for simplicity, which effectively reduces $S_7^{\rm cub, {\lambda}} $ to a term that contributes to the point-particle cubic-in-spin effect. The sum of these two terms are 
\begin{widetext}
\beq
\begin{aligned}
S_7^{{\rm cub}, pp}  + S_7^{\rm cub, {\lambda=1}}= & \left(- \frac{11667520}{3} \nu ^3 +\frac{158297798}{3}  \nu ^2 -\frac{143187259
   }{6} \nu + \frac{223835711}{24} \right)  \\[2pt]
& \times \frac{1442897}{\left( 113 - 76 \hskip 1pt \nu \right)^3 \left( 1444  \hskip 1pt \nu +8475 \right)} \hskip 2pt \chi^3 + \Delta S_7^{{\rm cub}, {pp}} \, , 
\end{aligned}
\eeq
where
\begin{align}
\Delta S_7^{{\rm cub}, pp} & = \frac{160 \left( 4308896 \nu +54927553 \right)}{\left( 113-76 \nu \right)^3} \delta  \nu ^3 \chi_a^3  + \frac{12769  \left( 2443362404 \nu +13380385211 \right)}{18606 \left( 76 \nu -113 \right)^3} \hskip 1pt  \delta \hskip 1pt   \nu \hskip 1pt  \chi_a \hskip 1pt  \chi^2 \nonumber \\ 
   & - \frac{674709440 \hskip 1pt  \pi  }{1329 \left( 113-76
   \nu \right)^2} \hskip 1pt  \delta \hskip 1pt  \nu \hskip 1pt  \chi_a \hskip 1pt  \chi + \frac{452  \left( 183323602308 \nu +10853245537 \right)}{3101 \left( 76 \nu -113 \right)^3} \hskip 1pt \nu ^2 \hskip 1pt  \chi_a^2 \hskip 1pt  \chi \, .
\end{align}
As before, we neglect the $\Delta S_7^{{\rm cub}, pp} $ term by taking $\chi_a = 0$. 
Finally, for the quadrupolar cubic-in-spin term $S_7^{\rm cub, {\kappa}} $, we utilize the fact that $\kappa_{\rm eff}^{\rm 2PN}$ and $\kappa_{\rm eff}^{\rm 3PN}$ scale quadratically in spins, and hence the only way in which cubic-in-spin terms could be obtained is by multiplying them by $\chi$ and $\chi_a$. 
We find
\beq
\begin{aligned}
S_7^{\rm cub, {\kappa}} & = \left( \frac{3110}{3} - \frac{73010 \hskip 1pt \nu}{113-76 \hskip 1pt \nu} \right) \kappa_{\rm eff}^{\rm 2PN} \chi -  \frac{448610 \hskip 1pt \nu^2}{3 (113-76 \hskip 1pt \nu)}  \kappa_{\rm eff}^{\rm 3PN} \chi + \Delta S_7^{\rm cub, {\kappa}} \, , \\[2pt]
 \Delta S_7^{\rm cub, {\kappa}} & = \frac{40 \hskip 1pt \nu \hskip 1pt \chi_a}{3 (113-76 \nu )} \sum_{i=1}^2 \sum_{j \neq i} (-1)^{i+1} \left( \frac{m_i}{m_1+m_2} \right)^2 \left(\frac{16691 m_i+20627 m_j}{m_1 + m_2} \right)\kappa_i \chi_i^2 \, ,
\end{aligned}
\eeq
\end{widetext}
where $\Delta S_7^{\rm cub, {\kappa}} \propto \chi_a$. 
Since we are largely interested in binary systems with large values of  $\kappa_i$, the truncation of $\Delta S_7^{\rm cub, {\kappa}}$ is the dominant source of loss in the effectualness of our reduced model.


To summarize, our reduced five-dimensional model consists of the reparameterizing of the dominant spin-induced quadrupolar phase terms with the effective quadrupole parameters $\kappa_{\rm eff}^{\rm 2PN}$ and $\kappa_{\rm eff}^{\rm 3PN}$. 
Furthermore, to achieve dimensional reduction, we express the phases as linear combinations of the $\chi$ and $\chi_a$ spin variables, and then truncate terms that depend on anti-symmetric spin variable by setting $\chi_a =0$. 
An effectualness study of this five-dimensional waveform will be presented in Sec.~\ref{sec:effectualness}.

\subsection{Reduced Four-Dimensional Waveform} \label{sec:4d}

The reduced 5D waveform presented in Sec.~\ref{sec:5d}, albeit more computationally managable than the original PN waveform, still has one more dimension than the waveform models used in standard compact binary searches. 
In this section, we discuss how we can further reduce the 5D waveform above to a 4D waveform that has comparatively little loss in effectualness. 
Note that four dimensions is the lowest possible number of dimensions for a waveform model that incorporates the spin-induced quadrupole as a free parameter, as the waveform must necessarily be described by the component masses, a reduced spin, and a single effective quadrupole parameter.


To arrive at a 4D waveform model, we retain the form of the five-dimensional model up until the 3PN terms, where the new parameter $\kappathree$ enters for the first time in \eqref{eqn:S6param}. 
Our goal is to re-express the spin-dependent 3PN and 3.5PN contributions to the waveform only in terms of the four parameters through the dimensional reduction $\textbf{p}_{5\mathrm{D}} = \{ m_1, m_2, \chi, \kappa_{\rm eff}^{\rm 2PN}, \kappa_{\rm eff}^{\rm 3PN} \} \to  \textbf{p}_{4\mathrm{D}} = \{ m_1, m_2, \chi, \kappa_{\rm eff}^{\rm 2PN}\} $. 
Since $\kappathree$ and $\kappatwo$ are linearly independent, this necessarily involves making a choice of combination of $\kappa_i$ parameters to truncate in our 4D model. 
To find a good choice, we focus on the $\kappa_i$-dependent part of $\kappathree$ in \eqref{eqn:kappa_eff_3PN} as it contributes to the largest phase deviation when $\kappa_i \gg 1$. 
We rewrite those terms as 
\begin{equation}
\begin{aligned}
\kappa_1 \chi_1^2 + \kappa_2 \chi_2^2  
= & g(\nu)\sum_i \left(\frac{m_i}{m}\right)^2 \kappa_i \chi_i^2
\\[4pt]
& 
+ \left[1 - \left( \frac{m_1}{m_2} \right)   g(\nu)  \hskip 1pt \nu \right] \kappa_1 \chi_1^2 
\\[4pt]
& 
+ \left[1 - \left( \frac{m_2}{m_1} \right) g(\nu) \hskip 1pt \nu \right] \kappa_2 \chi_2^2  \, , \label{eqn:kappa_3PN_tmp}
\end{aligned}
\end{equation}
where $g$ is \textit{a priori} an arbitrary function.
We have cast $\kappa_1 \chi_1^2 + \kappa_2 \chi_2^2$ in a form whereby the first term on the right-hand-side in \eqref{eqn:kappa_3PN_tmp} is proportional to the $\kappa_i$-dependent part of $\kappatwo$, cf.~\eqref{eqn:kappa_eff_2PN}.
 Our goal is to therefore find a suitable choice of $g$ such that the last two terms in \eqref{eqn:kappa_3PN_tmp} can be neglected with little impact on the phase.


\begin{figure}[t!]
\includegraphics[width=0.98\columnwidth]{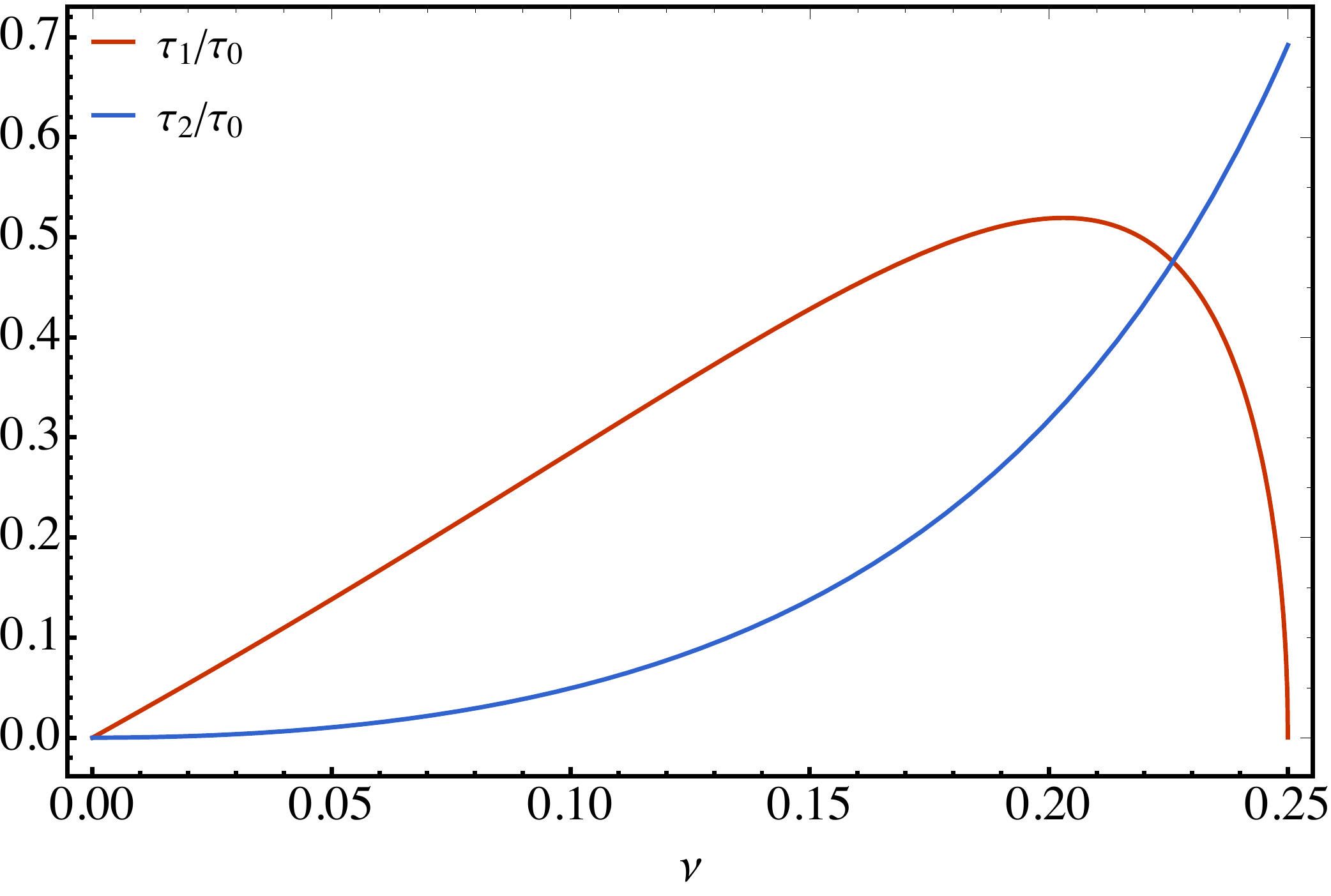} 
\caption{The mass-dependent prefactors arising in the 3PN terms in our reduced four-dimensional waveform. Our truncated model retains $\tau_0$, which is order unity, and neglects $\tau_1$ and $\tau_2$, which are smaller than $\tau_0$ for all mass ratios. }
\label{fig:tau}
\end{figure}

Our guiding principle in choosing $g$ is inspired by our truncation scheme at 2.5PN order, cf.~\eqref{eqn:S5truncation}. 
Recall that at that order, we truncate the two spin degrees of freedom to the single reduced spin parameter $\chi$ by neglecting terms that depend on the difference $\chi_a = (\chi_1 - \chi_2)/2$. 
Motivated by the success of this approach, we seek a truncation for \eqref{eqn:kappa_3PN_tmp} that only depends on the difference 
$\kappa_1 \chi_1^2 - \kappa_2 \chi_2^2$ and $\chi_a$. Enforcing this condition, we find $g = 2/(1-2\nu)$ and (\ref{eqn:kappa_eff_3PN}) becomes
\begin{equation}
\begin{aligned}
\kappathree  = & 
\frac{2 \kappatwo}{1-2 \nu}  + \frac{2}{1-2\nu}\left(
\tau_0 \chi^2 + \tau_1 \chi \chi_a + \tau_2 \chi_a^2
\right) \\[2pt]
& - \frac{\delta \left( \kappa_1 \chi_1^2 - \kappa_2 \chi_2^2 \right) }{1-2\nu}
 \\[2pt]
& - \frac{6}{2215 \nu^2} 
\left(\rho_2 \chi_a + \rho_3 \chi \chi_a + \rho_4 \chi_a^2 \right) \, ,
\end{aligned}
\end{equation}
where
\beq
\begin{aligned}
\tau_0 & = \frac{12769 \left( 250069 - 670112 \hskip 1pt \nu  + 57760 \hskip 1pt \nu ^2 \right) }
{20 \left( 113-76 \hskip 1pt  \nu \right)^2 \left( 8475 + 1444 \hskip 1pt  \nu \right)} \,,
\\
\tau_1 & = \frac{250069 }{5\left( 113-76 \hskip 1pt  \nu \right)^2}\hskip 1pt \delta \hskip 1pt \nu \,, 
\\
\tau_2 & = \frac{4\left( 82541+14440\hskip 1pt \nu \right)}{5\left( 113-76\hskip 1pt \nu \right) ^2} \hskip 1pt \nu^2 \, . 
\label{eqn:kappa3PN-kappa2PN}
\end{aligned}
\eeq
This reparameterization is advantageous in that the remaining $\kappa_i$-dependent terms vanish in the equal mass limit. 
The identity \eqref{eqn:kappa3PN-kappa2PN} leads us to a 4D model built from the 5D model by making the direct substitution
\begin{align}
\kappathree& \to \frac{2 \hskip 1pt  \kappatwo}{1-2 \hskip 1pt \nu}  + \frac{2\hskip 1pt  \tau_0 \chi^2}{1-2 \hskip 1pt  \nu} \, , \label{eqn:substitution}
\end{align}
which is equivalent to setting $\chi_a = 0$ as before, but also by ignoring the $\kappa_1 \chi_1^2 - \kappa_2 \chi^2_2$ term. 
We illustrate the mass-dependence of $\tau_1/\tau_0$ and $\tau_2/\tau_0$ in Fig.~\ref{fig:tau}, and find that the terms we neglect remain smaller than that of the term kept. 
Meanwhile, $\tau_0$ itself ranges within $0.70 \leq \tau_0 \leq 1.47$ as $\nu$ ranges between $0$ and $1/4$, so these spin terms make a small overall contribution to the phase. 
As we show in Sec.~\ref{sec:effectualness}, the difference in effectualness of the 4D waveform and the 5D waveform is surprisingly small.
For practical purposes we use the 4D waveform in our spin-induced quadrupole searches~\cite{Coogan:2022qxs, companion1} since there is little loss of effectualness combined with a substantial reduction in computational cost. 

\subsection{Waveform Cutoffs}
\label{sec:Cutoff}

In~\cite{Chia:2020psj}, the waveform for the inspiral of two exotic objects is cut off at a contact frequency which depends on the compactness $\mathcal C_i$ of the two bodies.
Ideally we would avoid introducing these two additional parameters, while still accounting for the fact that {\it a priori} we have no model for the merger and postmerger phase of the waveform.
A simple prescription would be to cut $\tilde h(f)$ at the ISCO of a Schwarzschild black hole with total mass $m$, 
$v_{\rm ISCO}^3 = \pi m f_{\rm ISCO} = 6^{-3/2}$, to remove the latter parts of the waveform from consideration.
However, it turns out that for systems with large spin-induced quadrupole moments, the binary must transition from an adiabatic inspiral into plunge and merger sooner than the black hole ISCO.
In some cases this occurs at a much lower frequency than the simple expectation. 

The adiabatic inspiral ends and the transition to plunge begins as the binary approaches the minimum of the binding energy $E(v)$, which determines the ISCO.
The spin-induced quadrupole moments contribute to the binding energy at 2PN and beyond.
The 2PN contribution has the opposite sign compared to the leading binding energy term, and moves the ISCO to larger radii and lower frequencies.
In fact, large $\kappa_i \chi_i^2$ values can dominate over the 1PN and 1.5PN corrections to $E$ when determining the minimum.
We use the minimum of $E$ as a starting point to calculate the frequency cutoff for our waveforms.

Unfortunately, using the binding energy minima alone to truncate the waveform still leads to problematic behaviour in the phase.
In particular, we find that $d\psi/df$ is not monotonic close to the truncation frequency, meaning that the time domain waveform obtained by performing a Fourier transform is also no longer monotonic (i.e. multiple frequencies contribute to the waveform at late times).
In principle, this should not occur: one can show using a convenient parametric solution of the phase~\cite{Tichy:1999pv} that $dE/dv = 0$ and $d^2\psi/df^2 = 0$ coincide.
In practice though, the resummation performed when computing the PN coefficients of the TaylorF2 model term by term introduces a slight shift in the minima.
We therefore multiply the frequency which minimizes the binding energy by an additional factor of $1/\sqrt{2}$.
We empirically find that this additional factor ensures that our frequency truncation is within 10\% of the frequency at which $d^2\psi/df^2 = 0$ whilst ensuring that $d\psi/df$ is always monotonic.

Our starting point then is the 3.5PN accurate expression for $E$ from~\cite{Kastha:2019brk}, which includes explicit $\kappa_i$-dependence up to quadratic order in spin, and we add the cubic in spin terms from~\cite{Marsat:2014xea}.
The binding energy has the form
\begin{align}
E(v) = -\frac{m \nu}{2} v^2 \sum_{n=0}^7 E_n(\textbf{p}_{6\mathrm{D}}) v^n \,,
\end{align}
where $E_0 = 1$, $E_1 = 0$, and we give the other coefficients in Appendix~\ref{sec:EnergyAppx}.

From here one option is to numerically solve $dE/dv=0$ to identify the minima. 
However, to ensure that $\tilde h(f)$ is both efficient to evaluate and differentiable,\footnote{In particular, this is to ensure that we can use the waveform to construct a template bank using the methods laid out in \cite{Coogan:2022qxs}.} we instead find an approximate series solution for the minima.
Since for large quadrupole moments $E_4 v^4$ can dominate over the lower order PN contributions at the ISCO, we create a baseline solution to iterate from by solving for the minimum of
\begin{align}
E_\kappa & =  -\frac{m \nu}{2} v^2(E_0 + E_4 v^4 + E_6 v^6) \,.
\end{align}
This yields a cubic equation with a single real root, $v^2 = v^2_0$, which can be written in closed from.

\begin{figure}[t!]
\includegraphics[width=0.98\columnwidth]{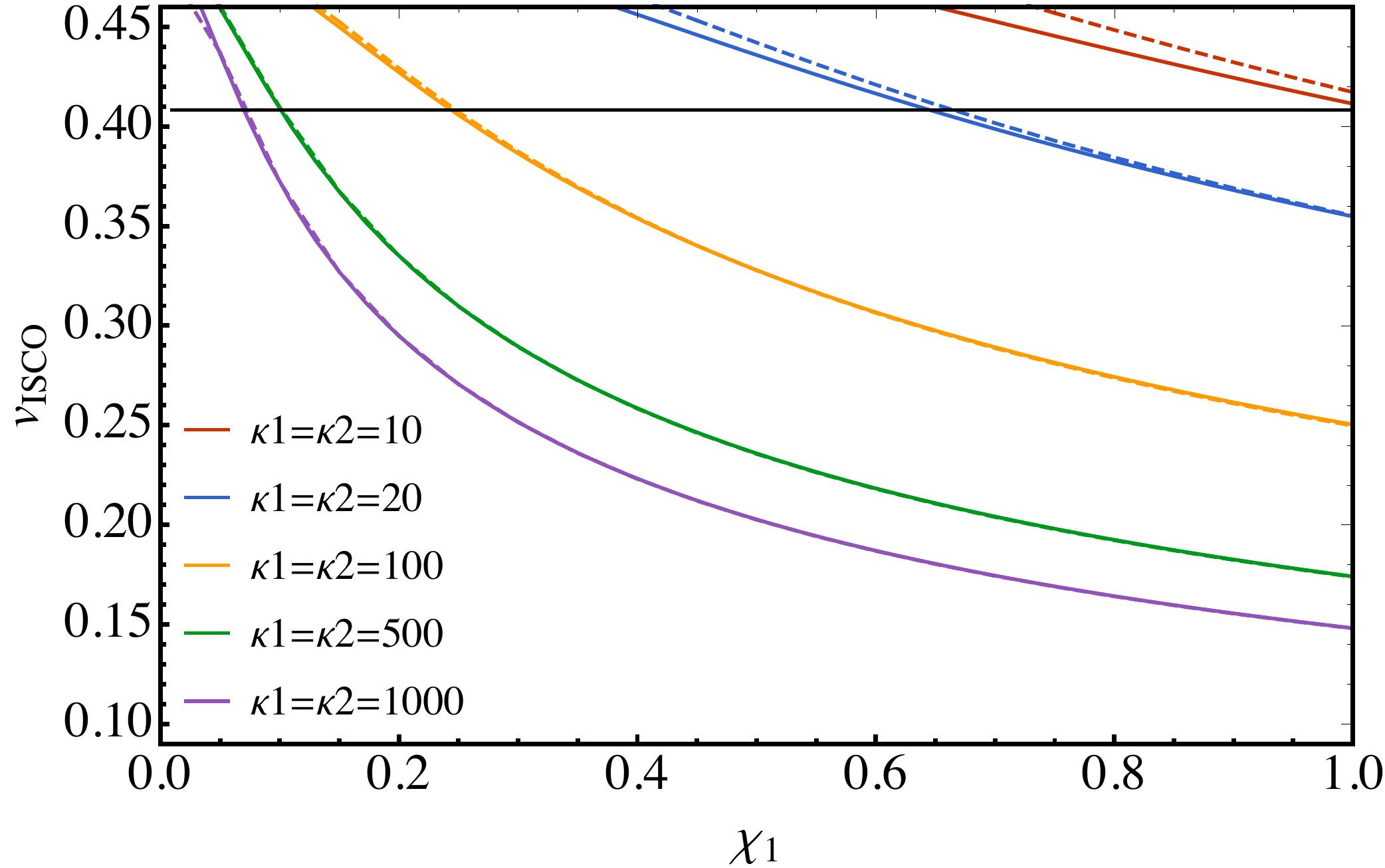} 
\caption{Comparison of the minimum of the 3.5PN binding energy $E(v)$ (dashed lines) to our analytic approximation to $v_{\rm ISCO}(\textbf{p}_{6\mathrm{D}})$ (solid lines) for $\nu = 0.2$, equal spins $\chi_1 = \chi_2$ and equal spin-induced quadrupole moments $\kappa_i$. The horizontal line marks $v_{\rm ISCO}$ for a Schwarzschild black hole of mass $m$.
}
\label{fig:ISCO}
\end{figure}

We introduce the remaining corrections to the energy order by order in the PN expansion, using a bookkeeping parameter $\epsilon$,
\begin{align}
E & = E_\kappa  -\frac{m \nu}{2} v^2 \left(\epsilon E_2 v^2 + \epsilon^2 E_3 v^3 + \epsilon^3 E_5 v^5\right)  \,.
\end{align}
We neglect the 3.5PN contribution $E_7$ in our analytic approximation.
In practice dropping this term has a negligible impact on the estimate.
To iterate our velocity solution, we let
\begin{align}
\label{eq:vISCO}
v_{\rm ISCO} = v_0 + \sum_{n=1}^5 \epsilon v_n\,,
\end{align}
so that we capture up to five orders of corrections in $v_{\rm ISCO}$,
and expand $dE/dv$ around $v_0$ in powers of $\epsilon$.
We solve this order by order in $\epsilon$, finally setting $\epsilon \to 1$.

Figure~\ref{fig:ISCO} illustrates the performance of our approximate, analytic estimate for $v_{\rm ISCO}$ compared to the numerical minumum of the 3.5PN energy for the 6D waveform.
Our approximation agrees well with the numerical solution for the minimum so long as the quadrupole contributions dominate $E$.
When this is not the case, as happens for small spins and for $\kappa_i$ near unity, the predicted $v_{\rm ISCO}$ is above that of the Schwarzschild black hole case, and so we use the lesser of the two values.
In practice, we also use the black hole $v_{\rm ISCO} = 6^{-1/2}$ when both $|\chi_{1,2} | < 0.05$ or $|\chi | < 0.025$.
Our approximation performs worst for anti-aligned spins, but remains close to the true ISCO and is generally conservative in the sense that it underestimates $v_{\rm ISCO}$ throughout the range of parameters we tested.

For the 5D and 4D models, we retain the same solution~\eqref{eq:vISCO} in terms of the coefficients of the binding energy $E_n$. 
To reduce the dimensionality of our approximation, we use truncated coefficients $E_n(\textbf{p}_{5\mathrm{D}})$ and $E_n(\textbf{p}_{4\mathrm{D}})$ for these models.
In particular, we follow the same strategy used previously to eliminate the individual $\chi_i$'s and $\kappa_i$'s in favor of $\chi$, $\kappa^{\rm 2PN}_{\rm eff}$, and $\kappa^{\rm 3PN}_{\rm eff}$. 
Our truncated coefficients $E_n$ are given for these models in Appendix~\ref{sec:EnergyAppx}.
Throughout the tested parameter space, we find that the ISCO predicted using these 5D and 4D energy coefficients is in very good agreement with the ISCO predicted using full $E_n(\textbf{p}_{6\mathrm{D}})$ expressions.

Finally, we note that the possibility that higher-order PN terms are boosted in importance due to large coefficients does not impact the assumptions behind the waveform model. 
One can show that the phase, which through the stationary phase approximation is built from PN expansions of $dE/dv$ and the flux $F(v)$~\cite{Tichy:1999pv}, takes the same form so long as the leading term in the PN series (i.e.~the Newtonian binding energy and the leading quadrupolar emission) dominates over the other PN corrections, which is of course the underlying assumption in the PN expansion.

\section{Effectualness studies}
\label{sec:effectualness}

We now want to characterize how the truncations made in the 5D and 4D reduced waveforms from the previous section translate into a loss in search efficiency.
Below we quantify the loss in signal-to-noise ratio (SNR) caused by using a 4D/5D waveform when trying to recover signals from the full 6D parameter space. More precisely, this loss is quantified via the \textit{effectualness} of the 4D/5D template to the full 6D inspiral waveforms.


We begin by defining the overlap
\begin{equation}
\label{eqn:overlap}
\left(h_{1} | h_{2}\right) \coloneqq 4 \operatorname{Re} \int_{f_l}^{f_u} \mathrm{d} f \frac{\tilde{h}_{1}(f) \tilde{h}_{2}^{*}(f)}{S_{\rm noise}(f)}\,,
\end{equation}
where $S_{\rm noise}$ is the one-sided noise spectral density and $\tilde{h}$ is the frequency domain waveform.\footnote{Throughout the rest of this work we use the frequency cutoffs $f_l = 12 \,\mathrm{Hz}$ and $f_u = 512 \,\mathrm{Hz}$. This is motivated by the fact that, for low-mass binary inspirals, $\gtrsim95$\% of the SNR is accumulated within this frequency range. Additionally, for all computations we use a frequency spacing of $\Delta f = 0.025 \, \mathrm{Hz}$.} 
Since we are interested in using these waveforms to search for signals in the O2 and O3 data releases, we use the $S_{\rm noise}$ presented in GWTC-2~\cite{LIGOScientific:2020ibl} from the Livingston detector (the most sensitive of the detectors).\footnote{\url{https://dcc.ligo.org/LIGO-P2000251/public}.} 
To simplify our notation, we additionally define the normalized overlap 
\begin{equation}
\label{eqn:normalized_overlap}
[h_1|h_2] \coloneqq (\hat{h}_1|\hat{h}_2) \, , \qquad \hat{h}_i \coloneqq h_i / (h_i|h_i)^{1/2}\,,
\end{equation}
where we have also defined the normalized signal waveform $\hat{h}_i$. 


\begin{figure*}
\centering
\includegraphics[width=1\linewidth]{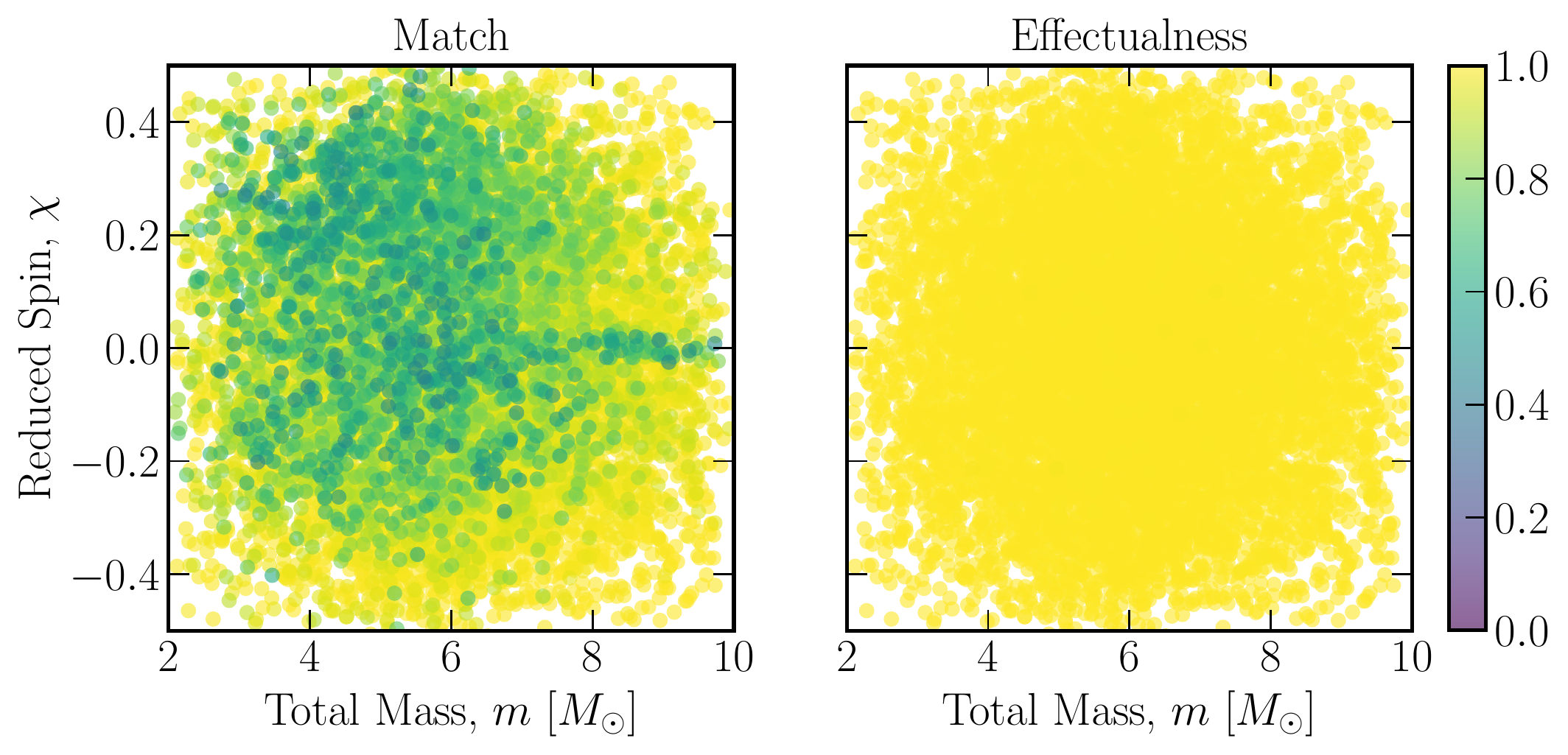}
\caption{(\textit{left}) The match (\ref{eqn:match}) between the 6D full inspiral waveform and the 4D truncated waveform for $10^4$ randomly sampled sets of parameters. (\textit{right}) The effectualness (\ref{eqn:effectual}) of the 4D waveform model to the same $10^4$ randomly sampled 6D full inspiral waveforms as before.  The left panel illustrates how our truncation scheme for the 4D model can introduce large errors in the phase, thereby significantly degrading the match between the original and the truncated models of the same parameters. However, the right panel shows that despite this significant reduction in the match, the 4D bank remains effective at capturing these 6D signals. Apart from the chirp mass, the inferred parameters of the best-fitting point in the template bank could be significantly different from the true source parameters of the signal.
}
\label{fig:matcheffectualness_chiredvstotalmass}
\end{figure*}

\begin{figure}
\centering
\includegraphics[width=1.0\linewidth ]{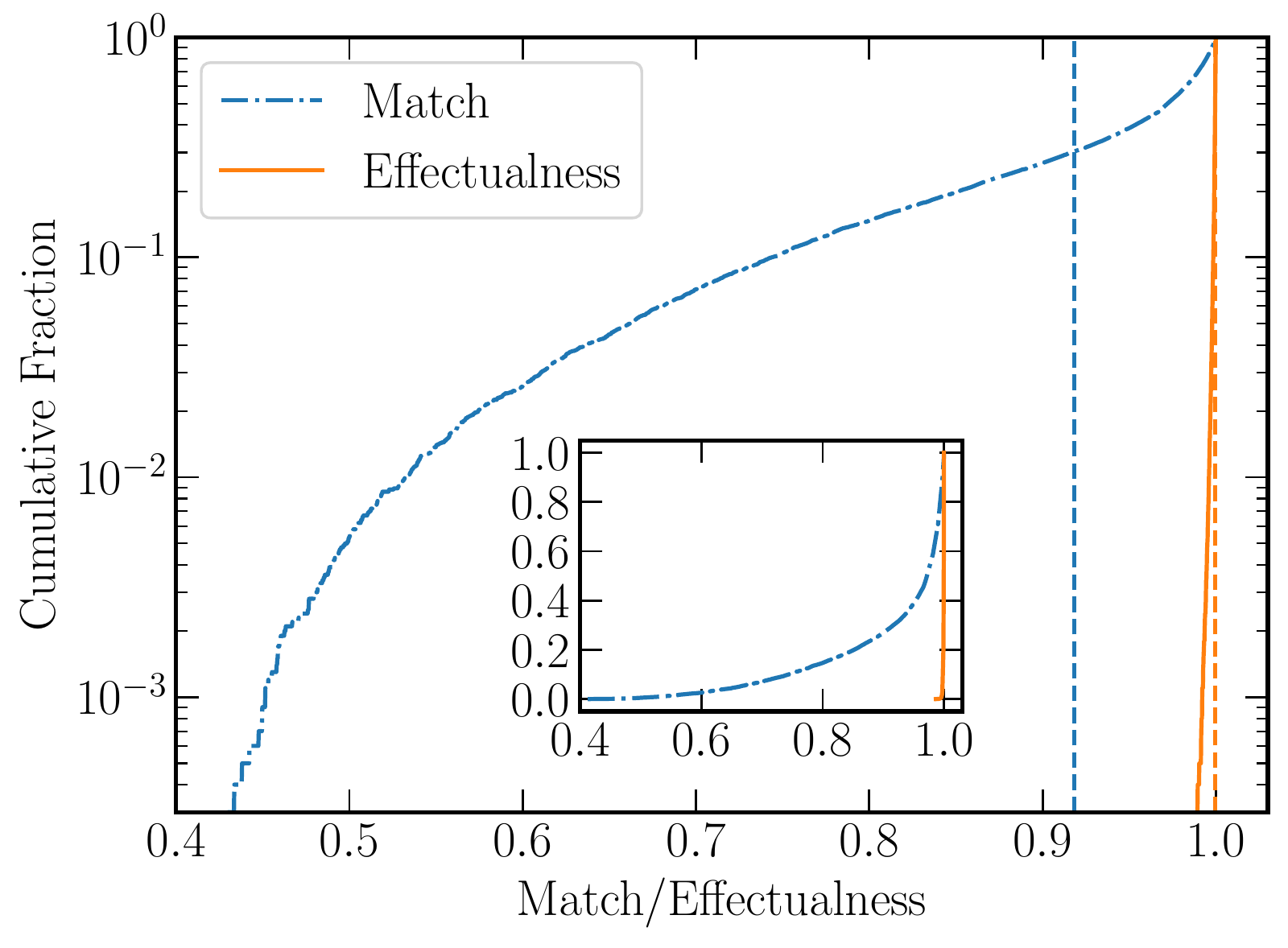}
\caption{Cumulative distribution functions of the match and effectualness results shown in Fig.~\ref{fig:matcheffectualness_chiredvstotalmass}. The vertical lines denote the mean of the distribution, which lie at $91.8\%$ and $99.9\%$ for the match and effectualness, respectively. The inset displays the same data on a linear-y-axis scale.
}
\label{fig:cdf_matchvseffectual}
\end{figure}

Since our primary aim in constructing the new waveforms in Sec.~\ref{sec:models} was to perform searches in a lower dimensional parameter space, our main quantity of interest is the reduction in SNR from using a 4D/5D waveform to search for signals whose waveforms are described by the full 6D parameter space. To this end, we introduce the notion of a \textit{match} between the waveforms with the 4D/5D parameters and the full 6D waveform:
\begin{align}
\label{eqn:match}
\mathcal{M}(\textbf{p}_{6\mathrm{D}},\textbf{p}_{4\mathrm{D}/5\mathrm{D}}) \coloneqq \underset{t_c,\phi_c}{\textrm{max}}  [h(\textbf{p}_{6\mathrm{D}})|h(\textbf{p}_{4\mathrm{D}/5\mathrm{D}})]\,,
\end{align}
which measures the distance between those waveforms.\footnote{Here, $t_c$ corresponds to the time of coalescence, and $\phi_c$ is the phase of coalescence. The precise definitions of $t_c$ and $\phi_c$ are not relevant to us since they are simply treated as extrinsic variables in the phase of the waveform, cf.~\eqref{eqn:waveform}. Maximizing over these parameters can therefore be easily accomplished by taking the absolute maximum of the Fourier transform of the inner product between the two waveforms given in (\ref{eqn:match}).} 
Note that our study from Sec.~\ref{sec:models}  focused on scenarios where the 4D/5D and 6D waveforms have the same intrinsic parameters, such that any mismatch between the 4D/5D and the full 6D waveform arises from our waveform truncation scheme and not a difference in intrinsic parameters.
The realized loss in SNR caused by using a 4D/5D waveform can then be quantified through the \textit{effectualness}:
\begin{align}
\label{eqn:effectual}
\varepsilon(\textbf{p}_{6\mathrm{D}}) \coloneqq \underset{t_c,\phi_c,\textbf{p}_{4\mathrm{D}/5\mathrm{D}}}{\textrm{max}}  [h(\textbf{p}_{6\mathrm{D}})|h(\textbf{p}_{4\mathrm{D}/5\mathrm{D}})]\,.
\end{align}
For our purposes, the effectualness describes how much of the SNR is retained when computing the overlap between a 6D waveform and a \textit{best-fitting} 4D/5D template waveform. 

Computing the effectualness requires one to perform a maximisation of the match over the 4D/5D parameter space for each given 6D signal waveform. 
Unfortunately, many maximizers fail to find the global best-fit point or are slow to converge, making it difficult to compute the effectualness across the full parameter space. 
We find that the differential evolution algorithm implemented in \texttt{scipy} \cite{storn1997differential} works well and converges sufficiently quickly to ensure that we are not computationally limited.

\begin{figure*}
\centering
\includegraphics[width=1\linewidth]{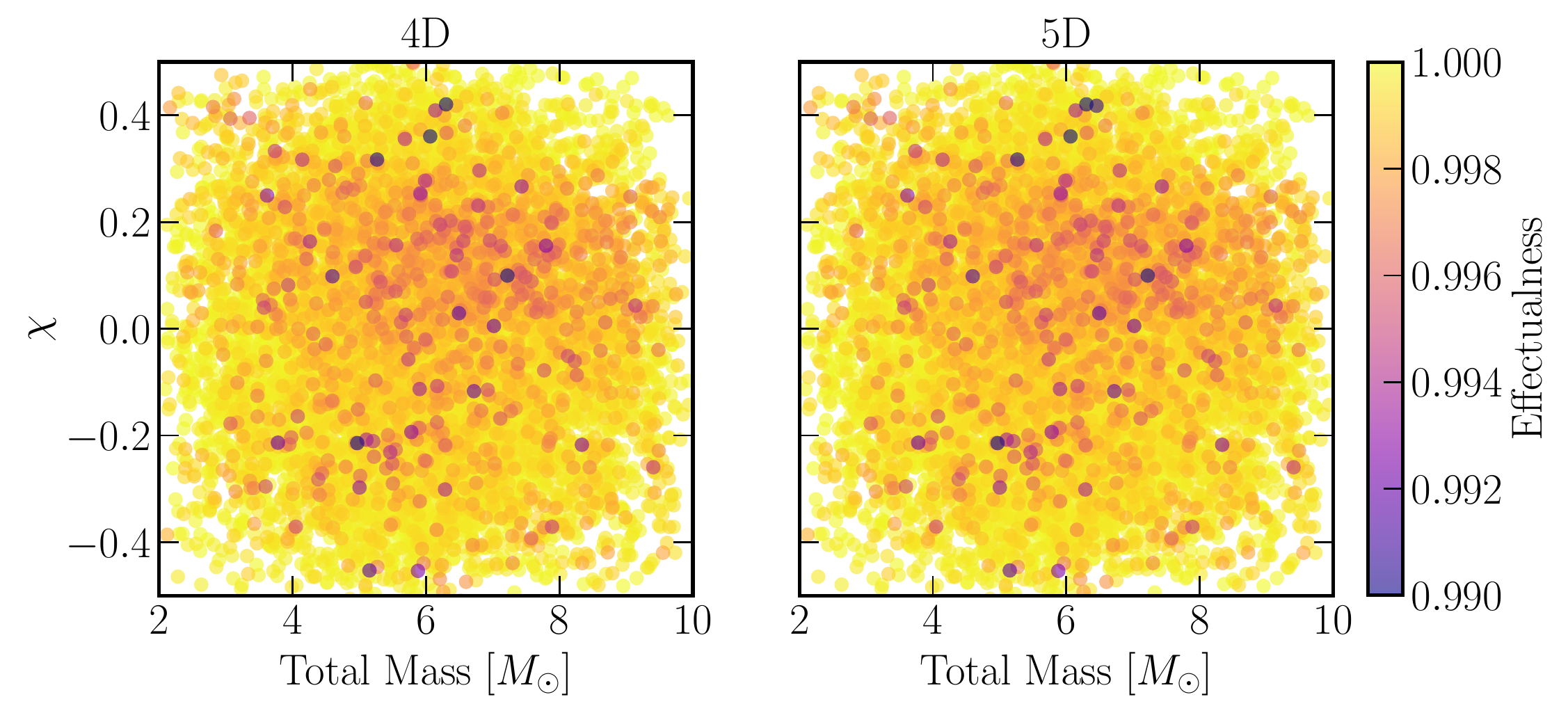}
\caption{Same as the right panel of Fig.~\ref{fig:matcheffectualness_chiredvstotalmass} except here we restrict the effectualness range to $[0.99, 1.0]$ and plot the results for both the 4D (\textit{left}) and 5D (\textit{right}) waveforms. 
We find that the effectualness is approximately independent of the total mass of the binary but does mildy depend on the reduced spin parameter $\chi$.
As expected, for $|\chi|\sim0$ the effectualness remains high whereas for $0.1\lesssim|\chi|\lesssim0.3$ there is some loss in effectualness.
Finally, for $|\chi| \gtrsim 0.3$, the effectualness starts to increase again (see text for discussion of possible explanation).
Note that the samples are ordered such that points with the lowest effectualness appear on top. 
}
\label{fig:effectualness_chiredvstotalmass}
\end{figure*}

\subsection{Match and Effectualness Studies} \label{sec:match_effectualness}

We are now able to quantify the loss in SNR incurred when searching for signals with the full 6D waveform using our dimensionally reduced models. 
Our results are shown in Fig.~\ref{fig:matcheffectualness_chiredvstotalmass}, where we inject $10^4$ 6D waveforms sampled uniformly from the parameter ranges: $1\,\mathrm{M}_\odot \leq m_{1,2} \leq 5\,\mathrm{M}_\odot$; $-0.6 \leq \chi_{1,2} \leq 0.6$, and $ 1 \leq \kappa_{1,2} \leq 1000$. 
In the left panel, we show the match (\ref{eqn:match}) between randomly injected waveforms and the equivalent 4D waveform with parameters found using (\ref{eqn:SO1.5}), (\ref{eqn:kappa_eff_2PN}), and (\ref{eqn:substitution}). 
The degradation of the match is a direct consequence of the fact that our truncation schemes in Sec.~\ref{sec:models} introduce large errors in the waveform phase. 
On the other hand, in the right panel, we find that the effectualness (\ref{eqn:effectual}) of our 4D template bank is remarkably high compared to the match across the majority of the parameter space.  
Note that both panels show the same set of $10^4$ injected points in the $\chi$ vs. $m=m_1 + m_2$ plane -- the 4D bank is therefore very effective at capturing the vast majority of signals with large spin-induced quadrupoles, although the best-fitting template could have significantly different parameters from the true signal source parameters. 
Indeed, we found empirically that although the injected and recovered chirp masses are always similar, the remaining parameters could differ significantly.\footnote{Note that for all scatter plots throughout this work we order the points such that the \textit{worst} points are plotted last. This visually emphasizes the points with the lowest match/effectualness.}

The difference in match and effectualness is more quantitatively illustrated in Fig.~\ref{fig:cdf_matchvseffectual}, where we show the cumulative distribution function of both quantities shown in Fig.~\ref{fig:matcheffectualness_chiredvstotalmass}.
In particular, the mean effectualness is found to be~99.9\% whereas the mean match is~91.8\%. Note however that the match has a long tail that extends to very low values of effectualness. In short, despite the low match between the 4D and 6D waveforms due to our truncation methods, the 4D template remains effective at capturing the full 6D inspiral signals. 

The high effectualness of our 4D waveform, shown in the right panel of Fig.~\ref{fig:matcheffectualness_chiredvstotalmass}, warrants deeper examination into the effectualness's dependencies on various source parameters. 
In Fig.~\ref{fig:effectualness_chiredvstotalmass}, we show the same results as the right panel of Fig.~\ref{fig:matcheffectualness_chiredvstotalmass}, except that we also show the results for the 5D model and reduce the range of effectualnesses to $\varepsilon \in [0.99, 1.0]$ in order to emphasize the differences across the parameter space. 
As expected, we see that both models perform well for $\chi\sim0$ since the effect of the spin-induced quadrupole is minimal, even for systems with large $\kappa$. 
For $|\chi|>0$, both models show a mild drop in effectualness as error introduced by our truncation in spins become increasingly important. 
Interestingly, for $|\chi|\gtrsim0.3$ this trend reverses and the effectualness starts to increase again.
This can be qualitatively understood through the effect of the truncations which reduce the total length of the waveform.
In particular, higher values of $\kappa_{1,2}$ combined with high spins ($|\chi_{1,2}|\gg0.0$ leads to $|\chi|\gtrsim0.3$) causes the 6D waveform to be truncated at a lower frequency, leaving a smaller number of cycles in band.
To maintain a high effectualness in this region of parameter space, our truncated waveforms therefore have to remain in phase with the 6D waveform for a smaller number of cycles.

\begin{figure*}
\centering
\includegraphics[width=1\linewidth]{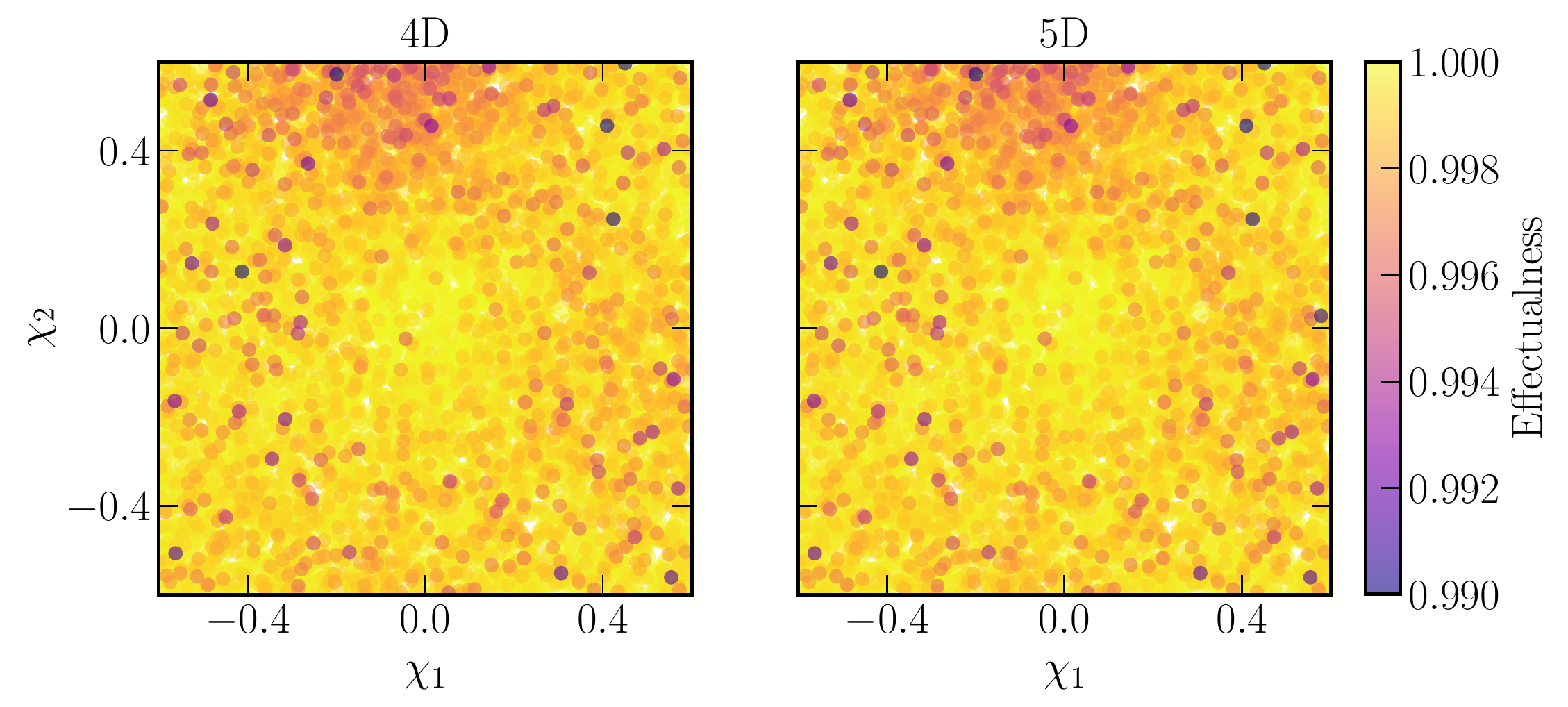}
\caption{Same as Fig.~\ref{fig:effectualness_chiredvstotalmass}, except here we plot the effectualness in the $\chi_1$ vs $\chi_2$ plane. We find that the effectualness is largely independent of the primary component spin, $\chi_1$, though it decreases fractionally as the spin of the secondary component exceeds $\chi_2 \gtrsim 0.4$.
}
\label{fig:effectualness_chis}
\end{figure*}

\begin{figure*}
\centering
\includegraphics[width=1\linewidth]{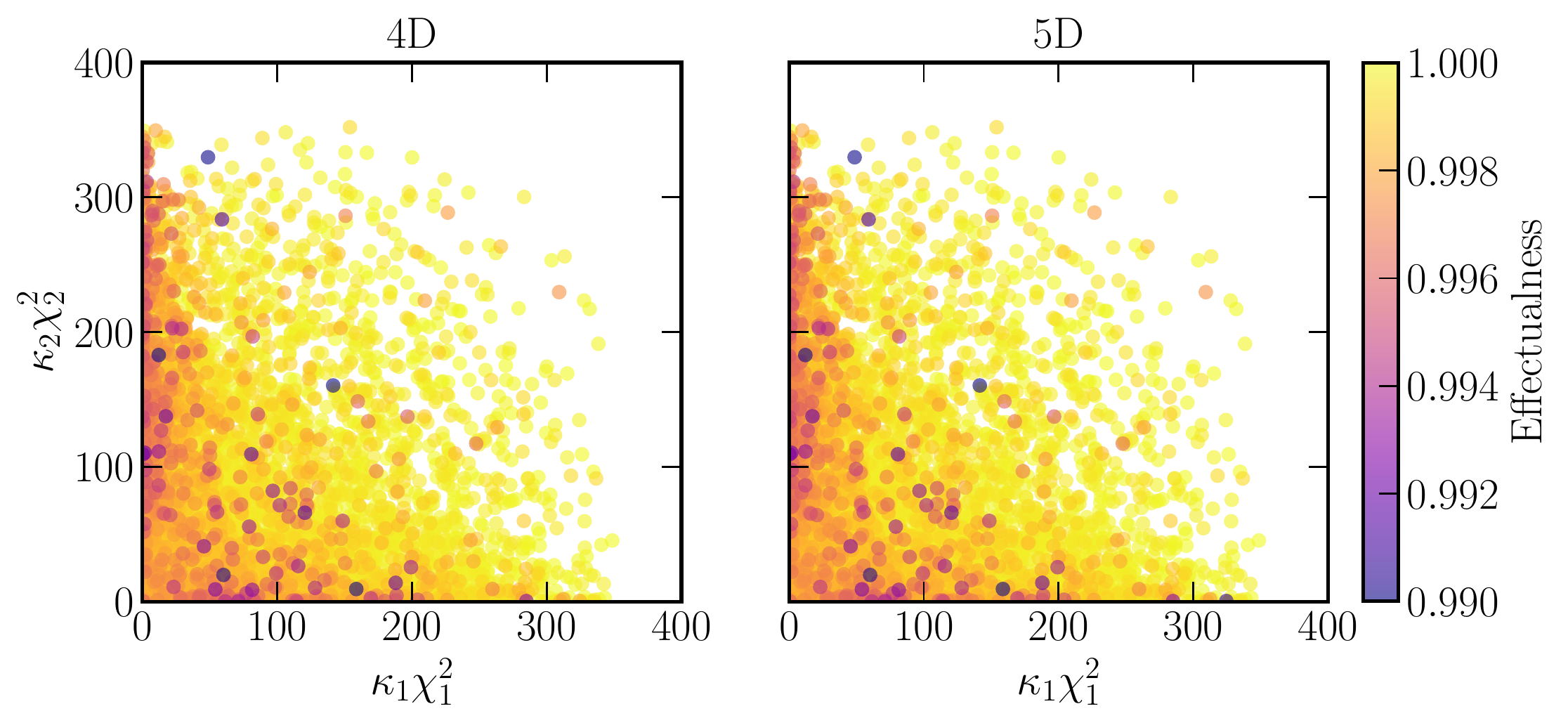}
\caption{Same as Figs.~\ref{fig:effectualness_chiredvstotalmass} and~\ref{fig:effectualness_chis}, except here we plot the effectualness as a function of $\kappa_1 \chi_1^2$ and $\kappa_2 \chi_2^2$, which are the parameter combinations that the spin-induced quadrupoles first appear in the phase, cf. Sec.~\ref{sec:models} and (\ref{eqn:kappa_def}). Similarly to Fig.~\ref{fig:effectualness_chis}, the effectualness decreases as $\chi_2$ and more importantly $\kappa_2$ increases. On the other hand, for the heavier binary component, the effectualness remains high even when its spin and quadrupole moments are large.
}
\label{fig:effectualness_chi2kappa}
\end{figure*}


\begin{figure}
\centering
\includegraphics[width=1.0\linewidth]{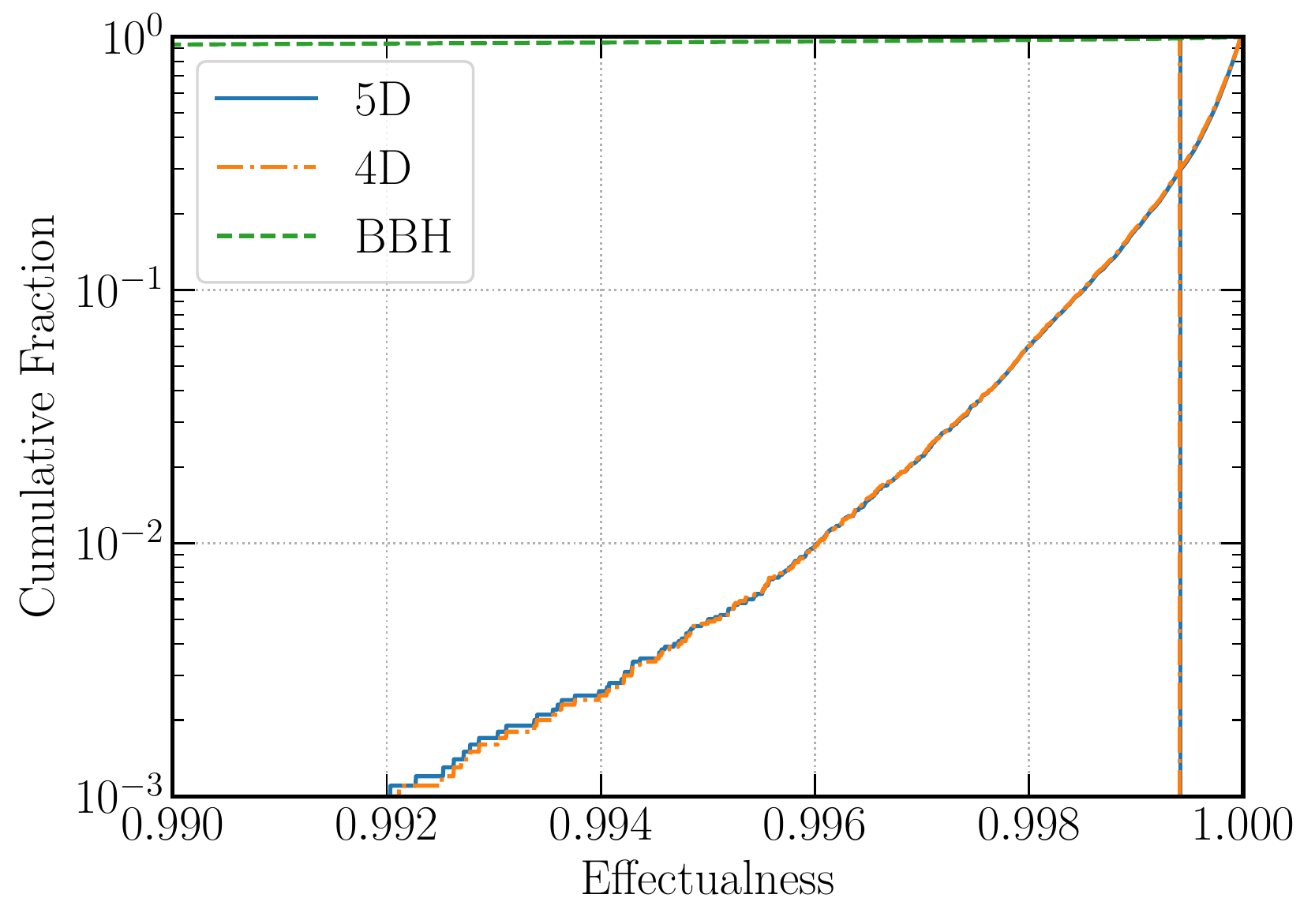}
\caption{ The cumulative distributions of the effectualness of our 4D and 5D template banks to $10^4$ injected 6D signal waveforms. Additionally, we show the effectualness of a BBH template bank to these same signal waveforms, illustrating how poorly BBH template banks that are used in existing search pipelines are at detecting these putative new signals. The mean effectualness of the 4D and 5D waveforms, as indicated by the vertical lines, are approximately $0.999$, while that of BBH waveform is $0.579$. 
}
\label{fig:cdf_effectualness}
\end{figure}

In Fig.~\ref{fig:effectualness_chis}, we show the effectualness of the same injected 6D waveforms as a function component spins, $\chi_1$ and $\chi_2$. 
For $\chi_{1,2}\sim0$, we find that both the 4D and 5D waveforms have high effectualness as expected. 
Interestingly, we observe that for $\chi_2 <0.4$ the effectualness is largely independent of $\chi_1$ and experiences no noticeable loss. 
For $\chi_2 >0.4$, there is some loss in effectualness which is particularly pronounced when $|\chi_1| <0.4$. 


Motivated by the parameterized spin-induced quadrupole relation (\ref{eqn:kappa_def}), we additionally plot the effectualness of the 4D/5D banks to the same $10^4$ 6D signal waveforms as a function of $\kappa_1\chi_1^2$ and $\kappa_2\chi_2^2$ in Fig.~\ref{fig:effectualness_chi2kappa}. 
Similarly to Fig.~\ref{fig:effectualness_chis}, we see the greatest loss in effectualness occurs for large values of $\kappa_2\chi_2^2 \gtrsim 100$.
As an instructive guide, we note that for $\chi_2=0.5$, the bound of $\kappa_2\chi_2^2 \leq 100$ corresponds to $\kappa_2 \leq 400$. 
Both of our reduced waveforms are therefore extremely effective at searching for binaries whose components have spin-induced quadrupoles that are large enough to evade detection by BBH template banks~\cite{Chia:2020psj}. 
Note that the effectualness degrades more appreciably with increasing $\kappa_2 \chi_2^2$  compared to increasing $\kappa_1 \chi_1^2$. 
This effect can be understood from the fact that the quadrupole of the heavier counterpart has a larger contribution to the binding energy, which shifts the binding energy's minimum towards a lower frequency and therefore truncates the waveform at an earlier stage of the inspiral; see \S\ref{sec:Cutoff} for our waveform cutoffs. 
A shorter waveform signal requires a smaller number of matching cycles between the 6D injection and the reduced waveforms for achieving a high effectualness, which explains why the effectualness in the large $\kappa_1 \chi_1^2$ region is higher than that when $\kappa_2 \chi_2^2$ is large.

A key takeaway of these injection studies is that the 4D and 5D waveform models have remarkably similar performance. 
This similarity is emphasized in Fig.~\ref{fig:cdf_effectualness}, where we show the cumulative distribution of the effectualnesses for both waveforms to the same $10^4$ signal waveforms used above.
For comparison, we also plot the effectualness of a standard BBH waveform model to the same signals, finding that it significantly underperforms our reduced models, as expected from earlier work~\cite{Chia:2020psj}.
The similarity between the models is evident with both waveforms showing agreement well into the tails of their distributions. 
Since in principle the 5D waveform is more flexible than the 4D model, its slightly worse effectualness is likely due to the minimizer's reduced performance in one additional dimension.
The similarity was somewhat unexpected, and suggests the existence of a non-trivial degeneracy between the 5D and 4D models. 
Fortunately, this also means that the 4D model can be used reliably in searches with no noticeable loss in effectualness across the same parameter space. 
Indeed, in a companion work, we use the 4D model to generate a template bank to perform the first search for binary waveforms with large spin-induced quadrupoles~\cite{Coogan:2022qxs, companion1}.

\subsection{Forecasted impact of modeling on constraints}

\begin{figure}[t!]
\centering
\includegraphics[width=1.0\linewidth]{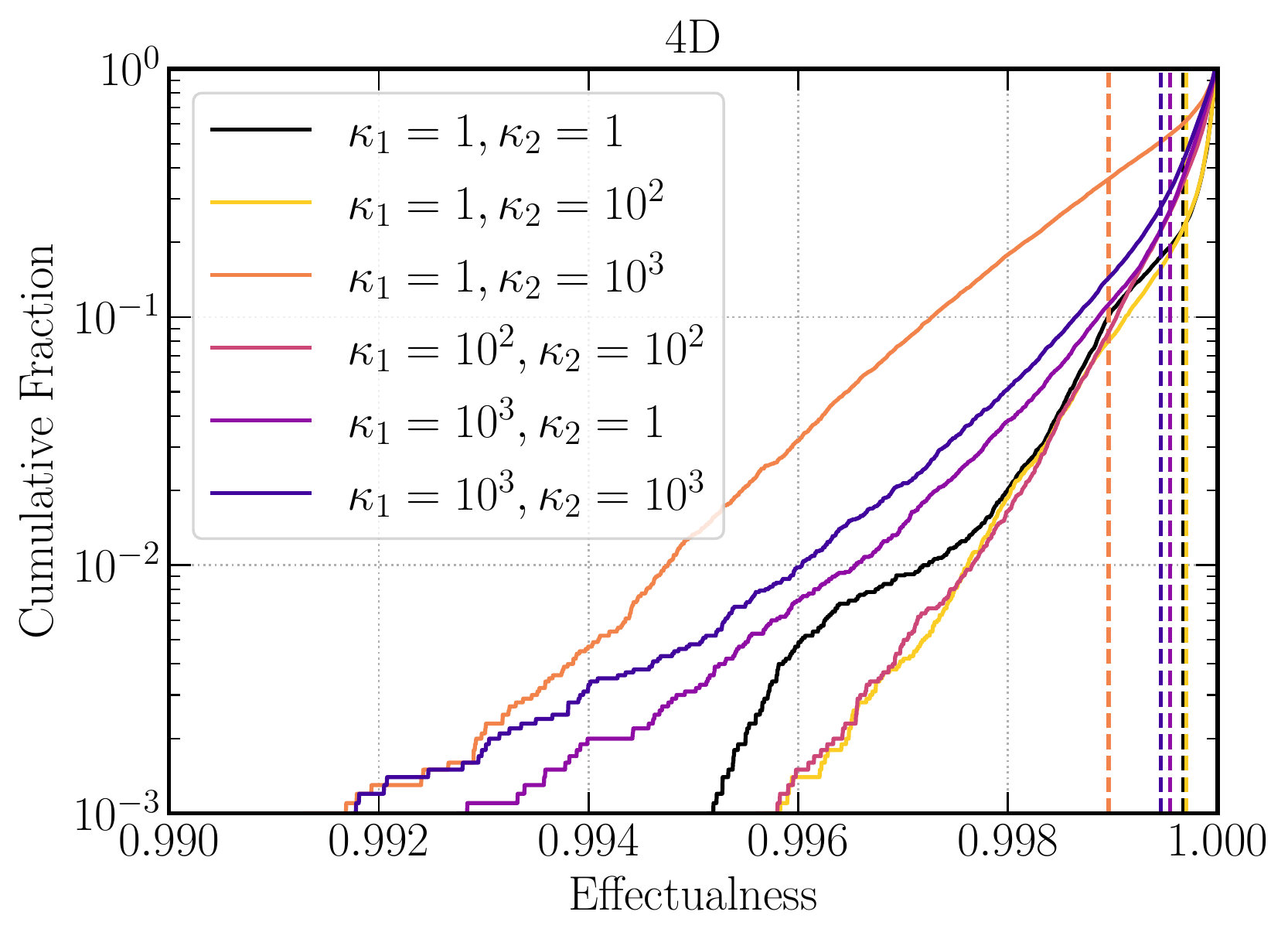}
\includegraphics[width=1.0\linewidth]{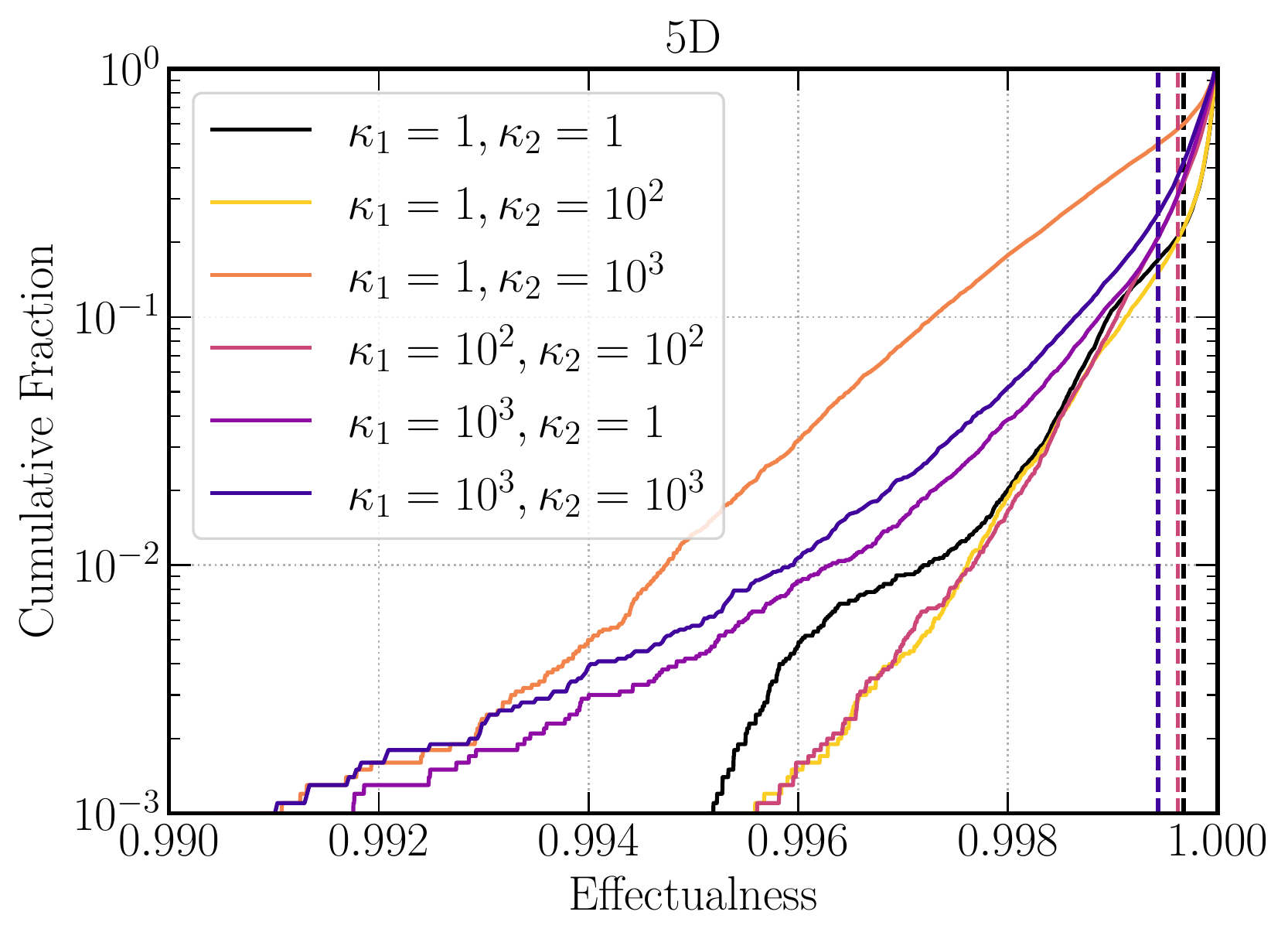}
\caption{Cumulative fractions from various injections sets along slices of the parameter space with fixed $\kappa_i$'s. Similarly to Fig.~\ref{fig:effectualness_chi2kappa}, we see that both $\kappa_1$ and $\kappa_2$ can take on any value within the tested range with little loss in effectualness. For larger values of $\kappa_{1,2}$, the mean effectualness (vertical lines) of the waveforms decrease, though 100\% of the injections still have an effectualness of $\varepsilon > 0.97$, which is the standard criterion demanded for BBH banks. As in previous figures, the 4D and 5D models perform similarly. Finally, for both models the $\kappa_1=\kappa_2=1$ injections are marginally worse than those with intermediate $\kappa$ values due to the boundaries imposed on the minimizer construction (see text for further discussion). }
\label{fig:cdf_fixedkappas}
\end{figure}

In this section, we further break down the performance of the truncated waveform models to specific values of $\kappa_i$'s. As described in \S\ref{sec:quad}, compact objects which are formed in various physics beyond the Standard Model scenarios, such as superradiant boson clouds and boson stars, have values of $\kappa_i$ that span over several orders of magnitude~\cite{Ryan:1996nk, Herdeiro:2014goa, Baumann:2018vus}. Here we perform injections tests with $10^4$ waveforms for a select set of combinations of $\kappa_1=\{1,100,1000\} $ and $\kappa_2=\{1,100,1000\}$, while the mass and spin ranges are sampled as above. Our goal is to allow the reader to more conveniently map the effectualness of our waveforms onto particular compact object model classes with specific $\kappa$ values.


In Fig.~\ref{fig:cdf_fixedkappas}, we show the cumulative distribution functions of our injection studies, where the trends are again similar between the 4D (left) and 5D (right) waveform models. 
We clearly see that $\kappa_1$ can vary up to $10^3$ without a significant drop in effectualness.
This behaviour was already seen along the top edge of Fig.~\ref{fig:effectualness_chis} and discussed in the surrounding text.
On the other hand, increasing $\kappa_2$ can reduce the overall effectualness of our truncated models.
Still, in the most extreme scenario we tested ($\kappa_1=\kappa_2=10^3$), the mean effectualness of the injections remains greater than $ \varepsilon = 0.999$. 
This clearly demonstrates that, even for binary systems with large spin-induced quadrupoles, our model is effective at detecting these putative new types of compact objects. 
Finally, we note that the $\kappa_1=\kappa_2=1$ injections show a marginally worse performance than those with intermediate $\kappa$ values.
We attribute this to the boundaries imposed on the differential evolution minimizer.
In particular, we demand that $\kappa^{\rm 2PN}_{\rm eff}$ remains positive, whereas in principle it can become slightly negative for unequal-mass binary systems (see \S\ref{sec:5d}).
Overall, we conclude that both our 4D and 5D waveform models maintain a high level of effectualness across all of the tested parameter space and are ideally suited for use in searches for compact objects with $\kappa_i \geq1$ up to relatively high spins $|\chi_{1,2}| \lesssim 0.6$.


\section{Conclusions and Outlook}
\label{sec:conclusions}

In this paper, we have developed 5D and 4D approximations to the 6D frequency domain waveforms used to describe the inspiral of two compact objects with arbitrarily large spin-induced quadrupoles~\cite{Krishnendu:2017shb}. Our dimensional-reduction method involves power counting in the PN expansion; suitable reparameterizations of the source physics in the waveform phase through the introcdution of the effective quadrupole parameters, $ \kappa_{\rm eff}^{\rm 2PN}$ and $ \kappa_{\rm eff}^{\rm 3PN}$ defined in (\ref{eqn:kappa_eff_2PN}) and  (\ref{eqn:S6param}); and truncating terms in the phase that depend explicitly on the antisymmetric spin, $\chi_a$. For the 4D model, we further truncate terms that depend on the combination $\kappa_1 \chi_1^2 - \kappa_2 \chi_2^2$, leading to the effective mapping from the 5D to 4D models through (\ref{eqn:substitution}).


Through a series of injection studies, we demonstrated that our reduced dimensional models are highly effectual to putative binary systems with large spin-induced quadrupoles across the majority of parameter space examined. 
In particular, we found that even systems in which $\kappa_{1,2}\lesssim10^3$, our truncated models lose only a marginal amount of effectualness across the parameter space. 
This can be seen clearly in Figs.~\ref{fig:effectualness_chi2kappa} and \ref{fig:cdf_fixedkappas}. 
Even the most extreme screnario we tested ($\kappa_1=\kappa_2=10^3$), our models still maintain a mean effectualness of 99.9\% across the $10^4$ injections. 
The high effectualness of our reduced models, despite the large degradation in the match shown in Fig.~\ref{fig:matcheffectualness_chiredvstotalmass}, indicates that our reduced waveforms can be reliably used to detect these new signals, though the inferred parameters would be different from the true source parameters. 
Indeed, we found empirically that although the chirp masses of the injected and recovered signals are always similar, other inferred parameters of the best-fitting point could be significantly different from the true parameters of the injected signal.


Interestingly, we find that there is little difference in the performance between the 4D and 5D models, despite naive expectations.
This suggests there is some none trivial degeneracy between the two models which should be explored further in future work.
Nevertheless, the similarity in performance between the two models suggests that one can use the 4D model to search for new signals with little loss in search performance.
This is particularly encouraging since this is the same number of dimensions used for typical BBH searches~\cite{LIGOScientific:2018mvr, LIGOScientific:2020ibl, LIGOScientific:2021djp}.


The primary purpose of finding these lower dimensional approximate waveforms was to enable searches without introducing onerous computational overheads. 
In companion papers, we use the 4D waveform developed here to construct a template bank (using the approach introduced in~\cite{Coogan:2022qxs}) and perform a search through the O3a data~\cite{companion1}. 
This is the first dedicated search for objects with large spin-induced quadrupole moments.
Future work should explore beyond large spin-induced quadrupoles to examine whether lower dimensional models can be used to capture a wide variety of binary systems.


Gravitational wave observations have provided a fundamentally new way to observe the Universe and look for signs of new physics. 
Since the most sensitive searches rely upon matched filtering to search for signals, it is therefore vital that waveforms are developed for a large variety of potential signals. 
A recent search using waveforms that include parameterized deviations from General Relativity~\cite{Narola:2022aob} has demonstrated how such an effort search could allow us to detect new physics.
We hope that this paper serves as a further guide to enable the community to go beyond BBH signals and ensure that searches for non-standard signatures are possible.

\acknowledgements
We are grateful to N.~.V~Krishnendu and K.~.G.~Arun for sharing {\it Mathematica} code associated with~\cite{Krishnendu:2017shb,Kastha:2019brk}. We thank David Trestini for helpful discussions on series expansions.
We especially thank Max Isi for reviewing a draft of this work and providing helpful feedback.
This research was supported in part by the National Science Foundation under Grant No.~NSF PHY-1748958. 
H.S.C.~gratefully acknowledges support from the Institute for Advanced Study and the Rubicon Fellowship awarded by the Netherlands Organisation for Scientific Research (NWO). 
T.E.~and K.F.~acknowledge support by the Vetenskapsr{\aa}det (Swedish Research Council) through contract No.~638-2013-8993 and the Oskar Klein Centre for Cosmoparticle Physics. 
T.E. is also supported by the Horizon Postdoctoral Fellowship.
R.N.G.~and A.~Z.~were supported by NSF grants PHY-1912578 and PHY-2207594. 
K.F.~is Jeff \& Gail Kodosky Endowed Chair in Physics at the University of Texas at Austin, and is grateful for support. 
K.F.~acknowledges funding from the U.S. Department of Energy, Office of Science, Office of High Energy Physics program under Award Number DE-SC0022021 at the University of Texas, Austin.
This research was done using resources provided by the Open Science Grid \cite{Pordes:2007zzb,Sfiligoi:2010zz}, which is supported by the National Science Foundation award \#2030508.

\begin{widetext}

\appendix 

\section{Details on the PN binding energy}
\label{sec:EnergyAppx}

Here we give the expressions for the PN binding energy coefficients $E_n$ discussed in Sec.~\ref{sec:Cutoff}, for our 4D, 5D, and 6D models.
We begin with the exact binding energy up to 3PN order (as discussed in the main text, for our approximation we take $E_7 \to 0$ for all our models), written both in a simple form and refactored to make the truncations simple,
\begin{align}
E_2 & = -\frac{3}{4} - \frac{\nu}{12} \,, \\
E_3 & = \frac{2}{3} \left[(2- \nu)(\chi_1 + \chi_2) + 2 \delta(\chi_1 - \chi_2)  \right] \,, \\
E_4 & = -\frac{27}{8} + \frac{19\nu}{8}  - \frac{\nu^2}{24} 
- \frac{1}{2}[( 1 + \delta - 2 \nu) \kappa_1 \chi_1^2 
+ (1 - \delta - 2 \nu) \kappa_2 \chi_2^2] 
- 2 \nu \chi_1 \chi_2 \,,  \\
E_5 & =  \frac{72 - 121 \nu + 2 \nu^2}{18}(\chi_1 + \chi_2) + \frac{72 - 31 \nu}{18}\delta(\chi_1 - \chi_2) \,, \\
E_6 & = - \frac{675}{64} 
+ \frac{(34 445 - 1230 \pi^2) \nu}{576} - \frac{155 \nu^2}{96} - \frac{35 \nu^3}{5184} 
\nonumber \\ & \hskip 10pt
+ \frac{5}{18}([4(1 + \delta) - \nu(1 - 7 \delta + 4 \nu)]\chi_1^2
+[4(1 - \delta) - \nu(1 + 7 \delta + 4 \nu)]\chi_2^2) -\frac{5\nu}{9} (3+\nu)\chi_1 \chi_2
\nonumber \\ & \hskip 10pt
-\frac{5}{12}
\left(
[7(1+\delta) - (19+ 5 \delta) \nu + 2 \nu^2]\kappa_1 \chi_1^2
+[7(1-\delta) - (19 - 5 \delta) \nu + 2 \nu^2]\kappa_2 \chi_2^2
\right)
\,.
\end{align}
The velocity expansion is given at leading order by
\begin{align}
v_0^2  &= - \frac{E_4}{4 E_6} 
+ \frac{1}{2} \left| \frac{E_4}{E_6}\right| 
\cosh \left[\frac{1}{3}\cosh^{-1} \left( -\frac{E_6(E_4^3 + 8 E_6^2)}{E_4^2 |E_4 E_6| }\right) \right] \,.
\end{align}
Note that analytic continuation is used for the $\cosh^{-1}$ when its argument leaves the domain of validity.
Regardless, $v_0^2$ remains real and positive.
The corrections to the ISCO velocity are
\begin{align}
v_1 & = -  \frac{E_2}{6 v_0 (E_4 +2 E_6 v_0^2)} \,, \\
v_2 & = - \frac{5 E_3 v_0^2 + 4 v_1(2 E_2 + 9 E_4 v_0 v_1 + 30 E_6 v_0^3 v_1)}{24v_0^2 (E_4 +2 E_6 v_0^2)} \,, \\
v_3 & =-  \frac{7 E_5 v_0^5 + 15 E_3 v_0^2 v_1 + 4 E_2(v_1^2 + 2 v_0 v_2)+24 E_4 v_0 v_1(v_1^2 + 3 v_0 v_2)+ 80 E_6 v_0^3 v_1(2v_1^2 + 3 v_0 v_2)}{24v_0^3 (E_4 +2 E_6 v_0^2)} \,, \\
v_4 & =  -\frac{1}{24v_0^3 (E_4 +2 E_6 v_0^2)}\left[
35 E_5 v_0^4 v_1 + 15 E_3 v_0(v_1^2 + v_0 v_2) + 8 E_2(v_1 v_2 + v_0 v_3) 
 \right.
\nonumber \\ & \hskip 10pt 
\left. 
+120 E_6 v_0^2(v_1^4 + 4 v_0 v_1^2 v_2 + v_0^2 v_2^2 + 2 v_0^2 v_1 v_3)
6 E_4 (v_1^4 + 12 v_0 v_1^2 v_2 + 6 v_0^2 v_2^2 + 12 v_0^2 v_1 v_3)
\right]\,, \\
v_5 & =  -\frac{1}{24v_0^3 (E_4 +2 E_6 v_0^2)}
\left[ 35 E_5 v_0^3 (2 v_1^2 +v_0 v_2) + 5 E_3(v_1^3 + 6 v_0 v_1 v_2 + 3 v_0^2 v_3)
+ 4 E_2(v_2^2 + 2 v_1 v_3 + 2 v_0 v_4)
 \right.
\nonumber \\ & \hskip 10pt 
\left. 
+24 E_4(v_1^3 v_2 + 3 v_0 v_1^2 v_3 + 3 v_0^2 v_2 v_3 + 3 v_0 v_1[v_2^2 + v_0 v_4])
 \right.
\nonumber \\ & \hskip 10pt 
\left. 
+48 E_6 v_0 (v_1^5 + 10 v_0 v_1^3 v_2 + 10 v_0^2 v_1^2 v_3 + 5 v_0^3 v_2 v_3 + 5 v_0^2v_1[2 v_2^2 + v_0 v_4])
\right]
\,.
\end{align}

For the 5D truncation, we refactor the energy coefficients in terms of $\textbf{p}_{5\mathrm{D}} = \{ m_1, m_2, \chi, \kappa_{\rm eff}^{\rm 2PN}, \kappa_{\rm eff}^{\rm 3PN} \}$ and residual parts that we truncate.
For example for $E_3$ we have
\begin{align}
E_3 & =  \frac{4}{3}\frac{113 (2 - \nu) \chi - 39 \nu \delta \chi_a}{113 - 76 \nu}\,,
\end{align}
and we set $\chi_a \to 0$ for our truncation after this factoring.
The forms we keep for the remaining coefficients are
\begin{align}
E_4 & \to  -\frac{27}{8} + \frac{19\nu}{8}  - \frac{\nu^2}{24} - \kappa^{\rm 2PN}_{\rm eff} 
- \frac{12769(250 069 - 8 \nu(41 389 - 14 440 \nu))}{20(113-76\nu)^2 (8475 + 1444 \nu)}\chi^2 \,,
\\
E_5 & \to \frac{113(72 - 121\nu + 2 \nu^2)}{9(113 - 76 \nu)} \chi \,, \\
E_6 & \to - \frac{675}{64} 
+ \frac{(34 445 - 1230 \pi^2 )\nu}{576} - \frac{155 \nu^2}{96} - \frac{35 \nu^3}{5184} 
- \frac{5(7 - 5 \nu)}{6}\kappa^{\rm 2PN}_{\rm eff} + \frac{10 \nu^2 }{3}\kappa^{\rm 2PN}_{\rm eff}
\nonumber \\ & \hskip 15pt
- 113^2 \frac{3 895 449 - 16 698 427 \nu + 13 190 680  \nu^2 - 577 600 \nu^3}{72(113 - 76 \nu)^2(8475 + 1444 \nu)} \chi^2 \,.
\end{align}
As shown in Fig.~\ref{fig:ISCO}, our analytic truncation provides a good approximation to the numeric result for $v_{\rm ISCO}$.

\end{widetext}

\bibliography{main}

\end{document}